\documentclass[11pt]{article}
\usepackage{cite}
\usepackage{scalerel}
\usepackage{amsmath,amsfonts,amssymb}
\usepackage[small,bf,hang]{caption}
\usepackage{slashed}
\usepackage{mathabx,amsmath}
\usepackage{latexsym,epsfig}
\usepackage{arydshln}
\usepackage{extarrows}

\usepackage{graphicx}
\usepackage{xcolor}
\usepackage{float}
\usepackage{tikz}
\usetikzlibrary{positioning,arrows,calc}
\usetikzlibrary{arrows.meta}
\usetikzlibrary{decorations.pathmorphing}
\usetikzlibrary{decorations.markings}
\usetikzlibrary{shapes.geometric}
\usetikzlibrary{patterns}

\newif\ifstartcompletesineup
\newif\ifendcompletesineup
\pgfkeys{
    /pgf/decoration/.cd,
    start up/.is if=startcompletesineup,
    start up=true,
    start up/.default=true,
    start down/.style={/pgf/decoration/start up=false},
    end up/.is if=endcompletesineup,
    end up=true,
    end up/.default=true,
    end down/.style={/pgf/decoration/end up=false}
}
\pgfdeclaredecoration{complete sines}{initial}
{
    \state{initial}[
        width=+0pt,
        next state=upsine,
        persistent precomputation={
            \ifstartcompletesineup
                \pgfkeys{/pgf/decoration automaton/next state=upsine}
                \ifendcompletesineup
                    \pgfmathsetmacro\matchinglength{
                        0.5*\pgfdecoratedinputsegmentlength / (ceil(0.5* \pgfdecoratedinputsegmentlength / \pgfdecorationsegmentlength) )
                    }
                \else
                    \pgfmathsetmacro\matchinglength{
                        0.5 * \pgfdecoratedinputsegmentlength / (ceil(0.5 * \pgfdecoratedinputsegmentlength / \pgfdecorationsegmentlength ) - 0.499)
                    }
                \fi
            \else
                \pgfkeys{/pgf/decoration automaton/next state=downsine}
                \ifendcompletesineup
                    \pgfmathsetmacro\matchinglength{
                        0.5* \pgfdecoratedinputsegmentlength / (ceil(0.5 * \pgfdecoratedinputsegmentlength / \pgfdecorationsegmentlength ) - 0.4999)
                    }
                \else
                    \pgfmathsetmacro\matchinglength{
                        0.5 * \pgfdecoratedinputsegmentlength / (ceil(0.5 * \pgfdecoratedinputsegmentlength / \pgfdecorationsegmentlength ) )
                    }
                \fi
            \fi
            \setlength{\pgfdecorationsegmentlength}{\matchinglength pt}
        }] {}
    \state{downsine}[width=\pgfdecorationsegmentlength,next state=upsine]{
        \pgfpathsine{\pgfpoint{0.5\pgfdecorationsegmentlength}{0.5\pgfdecorationsegmentamplitude}}
        \pgfpathcosine{\pgfpoint{0.5\pgfdecorationsegmentlength}{-0.5\pgfdecorationsegmentamplitude}}
    }
    \state{upsine}[width=\pgfdecorationsegmentlength,next state=downsine]{
        \pgfpathsine{\pgfpoint{0.5\pgfdecorationsegmentlength}{-0.5\pgfdecorationsegmentamplitude}}
        \pgfpathcosine{\pgfpoint{0.5\pgfdecorationsegmentlength}{0.5\pgfdecorationsegmentamplitude}}
}
    \state{final}{}
}

\tikzset{
corner/.style={line width=1pt,dashed,draw=black,dash pattern=on 6pt off 4pt},
scalar/.style={line width=1pt,draw=black},
photon/.style={line width=1pt,decorate, draw=black,
    decoration={complete sines,aspect=0,amplitude=1.25mm,segment length=1.5mm,start up,end up}},
}

\newcommand{\WEFF}{\begin{tikzpicture}[baseline={([yshift=-2ex]current bounding box.center)}]

\draw[photon] (0,1.5) -- (1,1.5);
\draw[fill=black] (1,1.5) circle (.05);
\draw[scalar] (1.5,1.5) circle (0.5);
\draw[fill=black] (2,1.5) circle (.05);
\draw[photon] (2,1.5) -- (3,1.5);

\node[] at (3.5,1.5) {\footnotesize $+$};

\draw[photon] (4,1.5) -- (6,1.5);
\draw[fill=black] (5.06,1.55) circle (.05);
\draw[scalar] (5.06,2.05) circle (.5);

\end{tikzpicture}}

\def\hybrid{
        \topmargin -20pt
        \oddsidemargin 0pt
        \headheight 0pt \headsep 0pt
        \textwidth 6.25in 
        \textheight 9.5in 
        \marginparwidth .875in
        \parskip 5pt plus 1pt \jot = 1.5ex}

\hybrid

 \csname
@addtoreset\endcsname{equation}{section}

\def\moth{\mathsurround=0pt}
\newdimen\zo \zo=0pt

\def\tick{\leaders\hrule height 0.5ex depth 0pt \hskip 0.5pt}
\def\upboxfill{$\moth \setbox\zo\hbox{\tick}%
  \hskip 3pt\hbox to 0pt{$\tick$\hss}\hrulefill \hbox to 7.5pt{$\tick$\hss}$}

\def\dtick{\leaders\hrule height .34pt depth 0.5ex \hskip 0.5pt}
\def\downboxfill{$\moth \setbox\zo\hbox{\dtick}%
  \hskip 2pt\hbox to 0pt{$\dtick$\hss}\hrulefill \hbox to 2pt{$\dtick$\hss}$}

\def\bec{\begin{center}}
\def\ec{\end{center}}
\def\a{\alpha}

\def\d{\delta}

\def\m{\mu}

\def\y{\eta}

\def\cB{{\cal B}}

\def\cJ{{\cal J}}
\def\cL{{\cal L}}
\def\cD{\mathfrak{D}}
\def\calD{\mathcal{D}}
\def\cF{{\cal F}}

\def\cA{{\cal A}}
\def\cB{{\cal B}}

\def\cN{{\cal N}}

\def\cH{{\cal H}}

\def\cA{{\cal A}}

\def\del{\partial}

\def\be{\begin{equation}}
\def\ee{\end{equation}}
\def\bea{\begin{eqnarray}}
\def\eea{\end{eqnarray}}
\def\ba{\begin{array}}
\def\ea{\end{array}}

\def\un{\underline}
\def\ov{\overline}

\begin{document}

\begin{titlepage}
\rightline{}
\rightline{August 2020}
\rightline{HU-EP-20/19}
\begin{center}
\vskip 1.5cm
  {\Large \bf{Old Dualities and New Anomalies}}
\vskip 1.7cm

{\large\bf {Roberto Bonezzi, Felipe D\'iaz-Jaramillo and Olaf Hohm}}
\vskip 1.6cm

{\it  Institute for Physics, Humboldt University Berlin,\\
 Zum Gro\ss en Windkanal 6, D-12489 Berlin, Germany}\\
 ohohm@physik.hu-berlin.de\\
roberto.bonezzi@physik.hu-berlin.de\\
felipe.diaz-jaramillo@hu-berlin.de
\vskip .1cm

\vskip .2cm

\end{center}

\bigskip\bigskip
\begin{center} 
\textbf{Abstract}

\end{center} 
\begin{quote}

We revisit the question whether the worldsheet theory of a string 
admits a global $O(d,d)$ symmetry. We consider the truncation of the target space theory in which   fields are independent of $d$ coordinates, 
which is  $O(d,d,\mathbb{R})$ invariant. The worldsheet theory is not $O(d,d,\mathbb{R})$ invariant, 
unless it is truncated by setting winding and center-of-mass momenta to zero. 
We prove consistency of this truncation and give a manifestly $O(d,d,\mathbb{R})$ invariant action,   
generalizing a formulation due to Tseytlin by including all external and internal target space fields. 
It is shown  that, due to chiral bosons, this symmetry is anomalous. The anomaly is cancelled 
by a Green-Schwarz mechanism that utilizes  the external B-field.

\end{quote} 
\vfill
\setcounter{footnote}{0}
\end{titlepage}

\tableofcontents


\section{Introduction}

T-duality is a property of string theory that emerges upon quantizing the string  on a toroidal background. 
Naturally, there have been numerous  papers addressing the question to which extent and in which sense 
the T-duality group $O(d,d)$, either in its discrete or continuous version,  is a duality or symmetry  of the (classical or quantum) worldsheet theory 
(see \cite{Tseytlin:1,Tseytlin:2,VENEZIANO1991287,Gasperini:1991ak,Kugo:1992md,Maharana:1992my,Siegel:1993th,Hull:2006va,Hull:2006qs,Berman:2007vi,Berman:2007xn,Berman:2007yf,Maharana:2010sp,Blair:2013noa,DeAngelis:2013wba,Hatsuda:2015cia,Bakas:2016nxt,Driezen:2016tnz} 
for an incomplete list of references). We nevertheless  come back to this issue, 
partly motivated by recent developments on the interplay of higher-derivative $\alpha'$ corrections and $O(d,d,\mathbb{R})$ invariance 
of the target space theory, both in conventional \cite{Eloy:2019hnl,Eloy:2020dko} and in double field theory 
formulations \cite{Hohm:2013jaa,Hohm:2014eba,Hohm:2014xsa,Marques:2015vua,Hohm:2016lge,Baron:2017dvb}. 
We will first identify a certain consistent truncation of the worldsheet theory of the bosonic string
in which $O(d,d,\mathbb{R})$ is a manifest symmetry classically, and second argue that this 
symmetry becomes anomalous quantum mechanically due to the worldsheet 
scalars being chiral bosons.
This in turn implies that  
a Green-Schwarz-type mechanism  is required in close  analogy to anomaly cancellation in heterotic string theory.

We begin by asking: 
Is the (classical or quantum) string worldsheet theory 
$O(d,d)$ invariant, either under the discrete or continuous group? 
The first relevant observation here is that in 
the Hamiltonian formulation of the worldsheet theory, for arbitrary backgrounds,  a `generalized metric' ${\cal H}_{MN}$ emerges that 
combines metric and B-field into an $O(d,d)$ matrix. However, as we will review, this does not mean that the worldsheet theory has a locally realized  $O(d,d)$ symmetry in general. 
A genuine $O(d,d)$ duality invariance is usually only expected to emerge on toroidal backgrounds. 
Suppose then that the target space is a torus in which a classical string propagates. Is the worldsheet theory $O(d,d)$ invariant? 
It is not, because there are winding modes that are discrete, due to the topology of the torus, while the center-of-mass momenta, which should pair up 
with the winding modes into an $O(d,d)$ multiplet,  are continuous. One could constrain the momenta by hand to be discrete, 
but there seems to be no physical justification for doing so. 
Similarly, one could promote the  winding numbers to dynamical fields (functions of worldsheet coordinates),
but then one is no longer dealing with a theory of strings since the worldsheet scalars are not well-defined maps on the torus preserving the torus   boundary conditions. 
Rather, the proper $O(d,d,\mathbb{Z})$ emerges when quantizing the worldsheet theory, because then the momenta are quantized, hence 
naturally pairing up with the discrete winding numbers. More precisely, the $O(d,d,\mathbb{Z})$ is then a duality (T-duality) in which a change of background 
leads to a physically equivalent theory in which momentum and winding is exchanged.

The above is standard text book folklore of string theory, but here we will revisit the issue from a slightly different point of view. 
We start from the observation that when one truncates the target space  theory by taking all fields to be independent of $d$ coordinates 
(for instance, by restricting to the massless fields  for Kaluza-Klein compactification on a torus $T^d$) a global $O(d,d,\mathbb{R})$ symmetry emerges.\footnote{This symmetry is in fact a consequence of the $O(d,d,\mathbb{Z})$ duality that target space closed string theory exhibits, since in the truncation 
both the usual massive Kaluza-Klein modes and their dual `winding' Kaluza-Klein modes disappear, so that the theory loses all memory of the torus 
topology, thereby enhancing $O(d,d,\mathbb{Z})$ to $O(d,d,\mathbb{R})$.}  If we now couple a classical string to this theory (in the same way 
that one may couple a point particle to Einstein gravity) does the combined system have an $O(d,d,\mathbb{R})$ symmetry? 
We will show that in general it does not, not even under the discrete subgroup, but that there is such a symmetry 
if one truncates also the worldsheet theory by setting winding and center-of-mass momenta to zero. 
This makes sense since it reflects  the truncation of the target space theory. Note that this truncation only refers to the internal sector 
(the coordinate directions on which the target space fields no longer depend), and so the string still has a non-trivial dynamics thanks to the external space. 
As one of our technical results we establish consistency of this truncation, for which the worldsheet action
takes  the manifestly $O(d,d,\mathbb{R})$ invariant 
form  
\begin{equation}\label{finalactionIntro}
\begin{split}
S=&-\frac{1}{4\pi\alpha'}\int d^2\sigma\Big[\sqrt{-h}\,h^{\alpha\beta} g_{\mu\nu} \del_\alpha X^\mu\del_\beta X^\nu
+\epsilon^{\alpha\beta}\big(B_{\mu\nu} \del_\alpha X^\mu\del_\beta X^\nu-\cA_\mu{}^M \partial_\alpha Y_M \del_\beta X^\mu\big)\Big]  \\
&+\frac{1}{4\pi\alpha'}\,\int d^2\sigma \Big[D_\sigma Y^MD_\tau Y_M-u\,D_\sigma Y^MD_\sigma Y_M-e\,\cH_{MN}\,D_\sigma Y^MD_\sigma Y^N\Big]\;, 
\end{split}
\end{equation}
where $X^{\mu}$ and $Y^M$ are the embedding scalars for the external and internal space, respectively, and 
$g_{\mu\nu}$, $B_{\mu\nu}$, ${\cal A}_{\mu}{}^{M}$ and ${\cal H}_{MN}$ are target space fields depending only on $X$. 
Moreover, we defined the covariant derivative 
\begin{equation}\label{DYIntro}
D_\alpha Y^M:=\del_\alpha Y^M+\cA_\mu{}^M(X)\del_\alpha X^\mu   \;,
\end{equation}
with worldsheet coordinates $\sigma^{\alpha}=(\tau,\sigma)$. 
This action is manifestly $O(d,d,\mathbb{R})$ invariant, with $M,N=1,\ldots, 2d$ being fundamental $O(d,d,\mathbb{R})$ indices. 
In particular, the internal scalars $Y^M$ are doubled, but the above action is equivalent to the standard sigma model action 
in this truncation, since the second order field equations imply, through integration and gauge fixing, first order duality relations. 
The above action generalizes a reformulation of the worldsheet action due to Tseytlin \cite{Tseytlin:1,Tseytlin:2} by including all 
external and internal target space fields that survive the truncation, in particular  the external B-field that turns 
out to be instrumental for the Green-Schwarz mechanism. The above Lagrangian, which provides an action principle for 
 equations of motion given by Maharana and Schwarz in \cite{Maharana:1992my}, 
 was also given by Schwarz and Sen in \cite{Schwarz:1993mg} and revisited recently in \cite{Blair:2016xnn}. 
 The action is indeed invariant under two-dimensional diffeomorphisms and Weyl transformations, albeit not manifestly so, 
 since the worldsheet coordinates have been split in the second line, where 
 $e$ and $u$ are the components of the 
 worldsheet metric $h_{\alpha\beta}$, defined by $e=\sqrt{-h}\,h_{\sigma\sigma}^{-1}$ and $u=h_{\tau\sigma}\,h_{\sigma\sigma}^{-1}$. 
We will give  a careful analysis of the non-manifest  two-dimensional 
diffeomorphism invariance and of the complete Virasoro constraints.

The results above apply to the classical worldsheet theory. 
As the second main point of this paper we then turn to its quantization and point out that since the worldsheet scalars are chiral (self-dual) 
bosons the $O(d,d,\mathbb{R})$ symmetry of the classical theory is expected to be anomalous. 
More precisely, it is technically and conceptually easier to work in a frame formulation, 
based on the coset space  $O(d,d,\mathbb{R})/SO(d,\mathbb{R})_L\times SO(d,\mathbb{R})_R$, where it is the gauge group $SO(d,\mathbb{R})_L\times SO(d,\mathbb{R})_R$
that becomes anomalous. 
The presence of anomalies  is confirmed independently 
by the recent result that in the target space theory the $O(d,d,\mathbb{R})$ symmetry, or alternatively the $SO(d,\mathbb{R})_L\times SO(d,\mathbb{R})_R$ symmetry,  
requires a deformation at order $\alpha'$ \cite{Eloy:2019hnl,Eloy:2020dko}, 
which from the point of view of the worldsheet theory cancels the anomaly 
via the Green-Schwarz mechanism. 
 This state of affairs mimics heterotic string theory, for which the worldsheet theory is anomalous due to the 
presence of chiral fermions, which gives a worldsheet interpretation of the Green-Schwarz mechanism \cite{Hull:1985jv,Sen:1985tq}. 
In the present context we prove that the one-loop effective action $W$ defined in terms of (\ref{finalactionIntro})  by 
 \be
  e^{iW[g,B,\cA, E]}=Z^{-1}\int DYe^{iS}
 \ee
transforms under $SO(d,\mathbb{R})_L\times SO(d,\mathbb{R})_R$ to lowest order as
 \be
  \delta_{\lambda}W =\frac{1}{8\pi}\int\,{\rm tr}\Big(d\lambda\wedge Q\Big)-\frac{1}{8\pi}\int\,{\rm tr}\Big(d\bar\lambda\wedge \bar Q\Big)\;.
 \ee
Here $E$ is a frame field for ${\cal H}_{MN}$ and $Q$ and $\bar{Q}$ are the (composite)  $SO(d,\mathbb{R})_L\times SO(d,\mathbb{R})_R$ connections. 
This anomaly is cancelled by assigning the following transformation to the B-field: 
 \be
  \delta_{\lambda} B_{\mu\nu}= \frac{\alpha'}{2}\,{\rm tr}\Big(\del_{[\mu}\lambda\, Q_{\nu]}\Big)-\frac{\alpha'}{2}\,{\rm tr}\Big(\del_{[\mu}\bar\lambda\,\bar Q_{\nu]}\Big)\;. 
 \ee
We also  
discuss and establish various other features regarding the 
quantum consistency of the worldsheet  theory, including  absence of gravitational anomalies \cite{AlvarezGaume:1983ig}, see also \cite{Roiban:2012gi,Hoare:2018jim}. 

This paper is organized as follows. 
In sec.~2 we review and clarify the Hamiltonian formulation of the worldsheet theory for strings with a toroidal target space. 
In particular, we introduce the proper truncation in which $O(d,d,\mathbb{R})$ will be made a manifest symmetry of the classical action. 
In sec.~3 we turn to the coupling of the worldsheet theory to the target space string theory and display the worldsheet action 
and equations of motion in a manifestly $O(d,d,\mathbb{R})$ invariant form. 
Then, in sec.~4, we show  the presence of anomalies and the need to 
invoke a Green-Schwarz mechanism. We close with a brief conclusion and outlook section. 
In two appendices  we perform a careful analysis of the diffemorphism invariance of the worldsheet theory, which is no longer manifest, 
and show that there are no gravitational anomalies.

\section{Classical String  on a Torus}

In this section we study the dynamics of a classical closed string on a toroidal target space $T^d\,$, representing the compact part of a $D=d+n$ dimensional spacetime with $d$ abelian isometries, 
and we shall focus for the moment on the dynamics along those directions alone.
In particular, we consider the string coordinate embeddings $X^i(\sigma,\tau)$ coupled to 
background fields $G_{ij}$ and $B_{ij}$ in the Polyakov sigma model
\begin{equation}\label{Polyakov constant}
S=-\frac{1}{4\pi\alpha'}\int d^2\sigma\Big[\sqrt{-h}h^{\alpha\beta}\del_\alpha X^i\del_\beta X^j\,G_{ij}+\epsilon^{\alpha\beta}\,\del_\alpha X^i\del_\beta X^j\,B_{ij}\Big]\;,  \end{equation}
where the worldsheet metric $h_{\alpha\beta}$ has Minkowski signature $(-,+)\,$, and $\epsilon^{01}=-1\,$. The compact space arises upon identifying $x^i\sim x^i+2\pi L^i$, where 
 $L^i:=\sqrt{\alpha'}w^i$ with integer winding numbers $w^i\in\mathbb{Z}\,$. Correspondingly, the allowed boundary conditions for the closed string worldsheet fields read (we use $\sigma\in[0,2\pi]$)
\begin{equation}\label{winding boundary conditions}
\Delta X^i(\tau):=X^i(2\pi, \tau)-X^i(0,\tau)=2\pi\,L^i\;,\quad \Delta h_{\alpha\beta}(\tau)=0\;.    
\end{equation}
The classical configuration space of the closed string is thus split by boundary conditions into the direct sum of disjoint topological sectors, labeled by the winding vector ${L}^i$. 
The non-compact case is covered by setting $L^i=0\,$. 
In view of the boundary conditions \eqref{winding boundary conditions} one can separate the winding sector 
as
\begin{equation}\label{splittingofwinding}
X^i(\sigma,\tau)=L^i\,\sigma+\overline{X}^i(\sigma,\tau)=L^i\,\sigma+\sum_{n\in\mathbb{Z}}x^i_n(\tau)\,e^{in\sigma}  \;,  
\end{equation}
since the shifted field obeys $\Delta\overline{X}^i=0\,.$ Let us mention that the variational principle with the action \eqref{Polyakov constant} is well defined, since neither $h_{\alpha\beta}\,$, $\del_\alpha X^i$ nor $\delta X^i$ have winding contributions.

\subsection{Hamiltonian Formulation}
We now turn to the Hamiltonian formulation, which turns out to be useful for identifying the symmetries. 
The  first step is to find the momenta conjugate to $X^i\,$:
\begin{equation}\label{compactPs}
P_i=\frac{1}{2\pi \alpha'}\big[\tfrac{1}{e}\,G_{ij}(\dot X^j-u\,X^{j\,'})+B_{ij}\,X^{j\,'}\big]\;,    
\end{equation}
where, as usual, a dot (prime) denotes a derivative w.r.t. $\tau\,(\sigma)\,$, and we defined 
the components of the worldsheet metric (that will become Hamiltonian Lagrange multipliers) via 
\begin{equation}\label{eu}
h_{\alpha\beta}=\frac{\Omega}{e}\,\begin{pmatrix}
u^2-e^2&u\\
u&1
\end{pmatrix} \;,\quad   h^{\alpha\beta}=\frac{1}{e\,\Omega}\,\begin{pmatrix}
-1&u\\u&e^2-u^2
\end{pmatrix}\,.
\end{equation}
The total Hamiltonian consists entirely of first class constraints, as it should be in any diffeomorphism  invariant theory. 
The action can then be written as 
\begin{equation}\label{Hamiltonaction}
S=\int d^2\sigma\,\Big[P_i\dot X^i-e\,\cH-u\,\cN\Big]\;, 
\end{equation}
where 
\begin{equation}\label{NH}
\begin{split}
\cN&:=P_iX^{i\,'}\;,\\
\cH&:=\frac12\Big[2\pi\alpha'\,G^{ij}P_iP_j-2\,G^{ik}B_{kj}\,P_iX^{j\,'}+\frac{1}{2\pi\alpha'}\,(G-BG^{-1}B)_{ij}\,X^{i\,'}X^{j\,'}\Big]  \,.
\end{split}
\end{equation}
Naturally, upon integrating out $P_i$ by solving its own equations of motion and back-substituting into the action one recovers 
the Polyakov action (\ref{Polyakov constant}). 
The functions in (\ref{NH}) are  phase space constraints (also referred to as Virasoro constraints) 
that 
are the canonical generators of worldsheet diffeomorphisms, being the Hamiltonian counterparts of the traceless worldsheet stress-energy tensor $T_{\alpha\beta}\,$. We notice that the third degree of freedom of the metric $h_{\alpha\beta}$, its overall conformal factor $\Omega\,$, drops out of the action.  In Hamiltonian language conformal gauge corresponds to gauge fixing $e=1$ and $u=0\,$.

Both Hamiltonian constraints can be put in a formally  $O(d,d)$ invariant form by defining
\begin{equation}\label{ZMvector}
Z^M:=\begin{pmatrix} \del_\sigma X^{i}\\2\pi\alpha'\,P_i
\end{pmatrix}\;,
\end{equation}
as well as the $O(d,d)$ invariant metric $\eta_{MN}$ and the generalized metric $\cH_{MN}$, 
\begin{equation}
\eta_{MN}=\begin{pmatrix} 0&\delta_i{}^j\\ \delta^i{}_j&0\end{pmatrix}   \;,\qquad \cH_{MN}=\begin{pmatrix} (G-BG^{-1}B)_{ij}&B_{ik}G^{kj}\\-G^{ik}B_{kj}&G^{ij}
\end{pmatrix}\;. 
\end{equation}
The functions in (\ref{NH}) can now be written as 
\begin{equation}\label{OddConstr}
\cN=\frac{1}{4\pi\alpha'}\,\eta_{MN}\,Z^MZ^N\;,\qquad \cH=\frac{1}{4\pi\alpha'}\,\cH_{MN}\,Z^M Z^N\;.    
\end{equation}
Although the Hamiltonian thus takes a formally $O(d,d)$ invariant form, the full action is not obviously $O(d,d)$ invariant. 
First, the $O(d,d)$ vector $Z^M$ in (\ref{ZMvector}) is defined in terms of derivatives of the fundamental field $X^i$ and hence 
it is not clear whether there is a locally realized  $O(d,d)$ symmetry even for the Hamiltonian terms. Second, the `symplectic current' $P_i\dot X^i$ is part of the action 
and not manifestly invariant.

 In order to elucidate the above issues, we will work explicitly with Fourier modes defined as
\begin{equation}
\begin{split}
X^i(\sigma,\tau)&=L^i\,\sigma+\sum_{n\in \mathbb{Z}}x^i_n(\tau)\,e^{in\sigma}\;,\quad P_i(\sigma,\tau)=\sum_{n\in \mathbb{Z}}p_{i\,n}(\tau)\,e^{in\sigma}\;,\\
e(\sigma,\tau)&=\sum_{n\in \mathbb{Z}}e_{n}(\tau)\,e^{in\sigma}\;,\qquad \qquad \;\, u(\sigma,\tau)=\sum_{n\in \mathbb{Z}}u_{n}(\tau)\,e^{in\sigma}\;,
\end{split}    
\end{equation}
where we recalled (\ref{splittingofwinding}). Here we assume 
the reality conditions $\varphi_n^*=\varphi_{-n}$ for all modes $\varphi_n:=(x^i_n, p_{i\,n}, e_n, u_n)\,$.
The action \eqref{Hamiltonaction} then reads 
\begin{equation}\label{Hamiltonactionmodes}
S=2\pi\,\int d\tau\sum_{n\in\mathbb{Z}}\Big[p_{i\,n}\dot x^i_{-n}-e_{-n}\,\cH_n-u_{-n}\, \cN_n\Big] \;.   
\end{equation}
Before giving the explicit expression for the modes $\cH_n$ and $\cN_n$ of the Virasoro constraints, let us introduce Fourier modes for the $O(d,d)$ vector $Z^M$ as
\begin{equation}\label{Zmodes}
\begin{split}
Z^M(\sigma,\tau)&=L^M(\tau)+\sum_{n\neq0}Z^M_n(\tau)\,e^{in\sigma}\;,\;{\rm where}\\
Z^M_n(\tau)&=\begin{pmatrix}in\,x^i_n(\tau)\\2\pi\alpha'p_{i\,n(\tau)}\end{pmatrix}\;,\;\; n\neq0\;,\qquad L^M(\tau)=\begin{pmatrix}L^i\\2\pi\alpha'p_{i\,0}(\tau)\end{pmatrix}    \;.
\end{split}
\end{equation}
At this point one can perform an invertible field redefinition by setting
\begin{equation}\label{Tseytlintrickmodes}
2\pi\alpha'\,p_{i\,n}(\tau)=:in\,\tilde x_{i\,n}(\tau)\;,\quad n\neq0  \;,
\end{equation}
that corresponds to defining dual coordinates $\widetilde  X_i(\sigma,\tau)$ via \cite{Tseytlin:1}
\begin{equation}\label{Tseytlintrick}
2\pi\alpha'\,P_i=\del_\sigma\widetilde X_i    \;.
\end{equation}
Integration of \eqref{Tseytlintrick} gives
\begin{equation}
\widetilde X_i(\sigma,\tau)=2\pi\alpha'\,p_{i\,0}(\tau)\sigma+\tilde x_{i\,0}(\tau)+\sum_{n\neq0}\tilde x_{i\,n}(\tau)\,e^{in\sigma} \;,   
\end{equation}
which introduces a zero mode $\tilde x_{i\,0}$ of $\widetilde X_i$ that does not appear in the original action, 
and for which it is not clear that it can become part of an $O(d,d)$ multiplet. 
Moreover, note that 
the dual fields $\widetilde X_i$ do not describe a closed string winding around a `dual torus', since their boundary conditions are not constant nor labeled by integers, 
and so in general the $\widetilde X_i$ cannot combine with the $X^i$ into an irreducible $O(d,d)$ representation. 
Nonetheless, from \eqref{Zmodes} and \eqref{Tseytlintrickmodes} one can still define the $O(d,d)$ non-zero modes
\begin{equation}
X^M_n(\tau):=\begin{pmatrix}x^i_n(\tau)\\\tilde x_{i\,n}(\tau)\end{pmatrix}    \;,\quad n\neq0\;,
\end{equation}
and $L^M(\tau)$ as in \eqref{Zmodes}.

Next, we rewrite the Virasoro constraints in terms of these Fourier modes. 
While the general Hamiltonian form of the action given above is valid for 
arbitrary backgrounds $G_{ij}$ and $B_{ij}$ 
we here focus on the torus and assume that 
the backgrounds are constant. 
The modes of the Virasoro constraints then take the formally $O(d,d)$ covariant form 
\begin{equation}
\begin{split}
&\cN_n=\frac{\eta_{MN}}{4\pi\alpha'}\big[2in\,X^M_nL^N-\sum_kk(n-k)X^M_kX^N_{n-k}\big]\;,\quad n\neq0\;,\\
&\cN_0=\frac{\eta_{MN}}{4\pi\alpha'}\big[L^ML^N+\sum_kk^2\,X^M_kX^N_{-k}\big]\,, \\
&\cH_n=\frac{\cH_{MN}}{4\pi\alpha'}\big[2in\,X^M_nL^N-\sum_kk(n-k)X^M_kX^N_{n-k}\big]\;,\quad n\neq0\;,\\
& \cH_0=\frac{\cH_{MN}}{4\pi\alpha'}\big[L^ML^N+\sum_kk^2\,X^M_kX^N_{-k}\big] \;,   
\end{split}    
\end{equation}
while the symplectic term can be recast in the form
\begin{equation}
2\pi\int d\tau\Big[p_{i\,0}\dot x^i_0+\frac{1}{4\pi\alpha'}\,\eta_{MN}\sum_{n\neq0}in\,X^M_{n}\dot X^N_{-n}\Big] \;. 
\end{equation}
Even though the non-zero modes $X^M_n$ can transform under an arbitrary $O(d,d)$ rotation as $X^{'M}_n=\Omega^M{}_N\,X^N_n\,$, the momentum-winding vector $L^M$ in general breaks the boundary conditions under $O(d,d)\,$: if we parametrize the $\Omega$ matrix as
\begin{equation}
\Omega^M{}_N=\begin{pmatrix}\omega^i{}_j & \alpha^{ij} \\
\beta_{ij} & \gamma_i{}^j\end{pmatrix}  \;,  
\end{equation}
 one has $L^{'i}=\omega^i{}_j\,L^j+2\pi\alpha'\,\alpha^{ij}\,p_{j\,0}(\tau)\,.$ The $\alpha$ transformation in particular yields a non acceptable boundary condition for the duality-rotated coordinate $X^{'i}\,$. Even if one is only interested in transforming classical solutions, where $p_{i\,0}(\tau)=k_i$ is constant, the discreteness of $L^i\,$, that descends purely from topology, is violated by $k_i\in \mathbb{R}^d$ for a general $O(d,d)$ rotation, even in the discrete subgroup $O(d,d,\mathbb{Z})\,$. 
As mentioned in the introduction,  one could truncate the spectrum by hand by taking the components $p_{i\,0}=k_i$ to be integers,  
thus mimicking the quantization condition that, however,  is not part of the original classical theory.
In contrast, at the quantum level the eigenvalues of the momentum zero mode $\hat p_{i\,0}$ take the discrete values $\frac{n_i}{2\pi\,\sqrt{\alpha'}}\,$, 
and so it is only here that one obtains the well-known T-duality group $O(d,d,\mathbb{Z})$.

It seems thus that neither the classical action nor the classical solutions of the closed string are invariant under $O(d,d)\,$. However, one can try to focus on a particular subsector of the classical theory: to begin with, we shall restrict the analysis to the topological sector with zero winding, \emph{i.e.}~${L^i=0}\,$. Classical solutions are also labeled by a constant center of mass momentum ${p_{i\,0}=k_i}\,$, and we will restrict to the class of zero momentum solutions, ${p_{i\,0}=0}\,$. Clearly, this subclass of solutions is closed under continuous $O(d,d,\mathbb{R})$ rotations, since $L^M=0$ is invariant.

Some comments are now in order: first of all, restricting to string solutions with vanishing center of mass momentum looks quite unphysical. However, one has to keep in mind that in the full theory (to which we shall turn in the next section) the compact space is only part of the entire spacetime. In fact, the low energy effective field theory, that displays the  $O(d,d,\mathbb{R})$ symmetry, precisely consists of fields that do not probe the internal space. In particular, they have zero Kaluza-Klein momenta and, obviously, cannot display winding. In this respect, it seems natural to consider the string dynamics restricted to zero winding \emph{and} internal momentum as the suitable probe for the low energy spacetime field theory.

\subsection{Truncated Dynamics and Consistency}

In order to restrict the space of classical solutions to zero winding and center of mass momentum, one can consider the truncated action obtained by setting $L^i=0$ and $p_{i\,0}=0$ in \eqref{Hamiltonactionmodes}:
\begin{equation}
S'=2\pi\int d\tau\sum_{n\in\mathbb{Z}}\Big[\frac{1}{4\pi\alpha'}\,in\,X^M_{n}\dot X_{-n\,M}-e_{-n}\,\cH_n-u_{-n}\,\cN_n\Big] \;, \end{equation}
with truncated Virasoro modes
\begin{equation}
\cN_n = -\frac{\eta_{MN}}{4\pi\alpha'}\sum_{k\in\mathbb{Z}}k(n-k)X^M_kX^N_{n-k}\;,\quad \cH_n = -\frac{\cH_{MN}}{4\pi\alpha'}\sum_{k\in\mathbb{Z}}k(n-k)X^M_kX^N_{n-k}\;.
\end{equation}
The above action, which is manifestly $O(d,d,\mathbb{R})$ invariant, can be rewritten in local form as
\begin{equation}\label{Tseytlincompact}
S'=\frac{1}{4\pi\alpha'}\,\int d^2\sigma\Big[\del_\sigma X^M\del_\tau X_M-e\,\cH_{MN}\,\del_\sigma X^M\del_\sigma X^N-u\,\del_\sigma X^M\del_\sigma X_M\Big]\;. \end{equation}
This is Tseytlin's original proposal, but with two-dimensional diffeomorphism invariance left intact (albeit in a non-manifest form). 
Taking $X^M(2\pi,\tau)=X^M(0,\tau)$ automatically sets both winding and center of mass momentum to zero, since $2\pi\alpha'\,p_{i\,0}=\tfrac{1}{2\pi}\,\int_0^{2\pi}d\sigma\,\del_\sigma\widetilde X_i=0\,$.

The equations of motion for the non-zero modes $X^M_n$ (or, which is the same, $x^i_n$ and $p_{i\,n}$) as well as the Virasoro constraints coincide with the original ones obtained from \eqref{Hamiltonaction} or \eqref{Hamiltonactionmodes} upon choosing the solution $p_{i\,0}=0\,$. On the other hand, one has to be more careful with the zero modes $X^M_0\,$: neither $x^i_0$ nor $\tilde x_{i\,0}$ appear in the action \eqref{Tseytlincompact}, that indeed has the obvious gauge symmetry $\delta X^M=\Xi^M(\tau)\,$. While this is fine for $\tilde x_{i\,0}\,$, for which it is just a redundancy of the field redefinition $2\pi\alpha'\,P_i=\del_\sigma\widetilde X_i\,$, it is \emph{not} equivalent for $x^i_0$ that does possess a non-trivial equation of motion in the original theory. We view  the reduced action \eqref{Tseytlincompact} as providing the dynamics for the non-zero modes $X^M_n\,$ 
and then establish that this is a consistent truncation of the full worldsheet theory. To this end we have to show that 
once a solution is provided for the non-zero modes (modulo worldsheet diffeomorphisms), we can embed it into a solution of the full theory. 
This means that we have to give $x^i_0$ in terms of the untruncated  fields so that the complete equations of motion of the original theory are satisfied.

Let us then study the original field equation for the zero mode $x^i_0(\tau)$.
Generally, the second-order Lagrangian equations of motion are  equivalent to the two sets of Hamiltionian equations obtained by varying with respect to $P_i$ and $X^i$, respectively. 
The former equation can be obtained by inverting the definition of canonical momenta \eqref{compactPs}, 
 \be\label{dotXiup}
  \dot{X}^i = u\,X^{j\prime}+ e\,G^{ij}\big(2\pi\alpha' P_j-B_{jk} X^{k\prime}\big)\,. 
 \ee 
The equation for the zero mode $x^i_0(\tau)$ can then be obtained by integrating over $\sigma$, 
\begin{equation}\label{x0equation}
\dot x^i_0(\tau)=V^i(\tau)\;,    
\end{equation}
where from (\ref{dotXiup}) we notice that $V^i$ is naturally the  upper component of the $O(d,d)$ vector
\begin{equation}\label{upperandlowercomponents}
V^M(\tau)=\begin{pmatrix}
V^i(\tau)\\\widetilde V_i(\tau)
\end{pmatrix}:=\frac{1}{2\pi}\,\int_0^{2\pi}d\sigma\big[u\,\del_\sigma X^M+e\,\cH^{MN}\,\del_\sigma
 X_N\big]   \;. 
\end{equation}
The Hamiltonian equation obtained by varying with respect to $X^i$ reduces for the zero-modes to $\dot{p}_{i\,0}=0$, since 
the functions (\ref{NH}) are independent of $x_0^i$ (they depend only on $X^{i\prime}$). Thus, these equations are trivially satisfied 
for $p_{i\,0}=0$, and so we only have to worry about eq.~(\ref{x0equation}). 
Given a solution to the field equations derived from \eqref{Tseytlincompact} (that leave the zero modes completely undetermined), one can directly integrate \eqref{x0equation}:
\begin{equation}\label{x0solution}
x^i_0(\tau)= x^i+
\int_0^\tau d\tau'\,V^i(\tau')\;. 
\end{equation}
This is the embedding into the full theory, which by construction satisfies the equations of motion. 
Note that we could use the lower component of (\ref{upperandlowercomponents}) to similarly define a function $\tilde x_{i\,0}(\tau)$, 
but there is no need to do so since such a dynamical variable does not appear in the original  theory.

\section{General Worldsheet Action }

\subsection{Kaluza-Klein Split}

In this section we are going to consider the more general sigma model of a closed string propagating on a target $(D=d+n)$-dimensional spacetime characterized by $d$ abelian isometries. We will choose coordinates $\hat x^{\hat\mu}=(x^\mu, y^i)\,$, with $\mu=0,...,n-1$ and $i=1,...,d$ such that all spacetime fields are independent of $y^i\,$, being the isometry directions, either compact or not. 
The  $n$-dimensional spacetime field content consists of \cite{Maharana:1992my}:  
\begin{itemize}
\item the $n$-dimensional metric, dilaton and Kalb-Ramond fields $g_{\mu\nu}\,$, $\phi$ and $B_{\mu\nu}\,$,\\
\item $2d$ abelian gauge fields forming an $O(d,d)$ vector: 
\be
 \cA_\mu{}^M=\begin{pmatrix}
 A_\mu^i\\\widetilde A_{\mu\,i}
\end{pmatrix}\,
\ee
which originate  from the off-diagonal components of the higher dimensional metric $\hat G_{\hat\mu\hat\nu}$ and $B$-field $\hat B_{\hat\mu\hat\nu}\,$,\\
\item $d^2$ scalar fields $G_{ij}$ and $B_{ij}$ originating  from the internal components of $\hat G_{\hat\mu\hat\nu}$ and $\hat B_{\hat\mu\hat\nu}\,$, that organize into the $O(d,d)$ valued generalized metric $\cH_{MN}$.
\end{itemize}

The reduced $n$-dimensional effective field theory action reads \cite{Maharana:1992my}
\begin{equation}\label{LEEA}
\begin{split}
S_{\rm FT}=\frac{1}{2\kappa^2}\int d^nx\sqrt{-g}\,e^{-2\phi}\Big[&R+4\,\del_\mu\phi\,\del^\mu\phi-\tfrac{1}{12}\,H_{\mu\nu\rho}\,H^{\mu\nu\rho}\\
&-\tfrac14\,\cH_{MN}\,\cF_{\mu\nu}{}^M\,\cF^{\mu\nu\,N}+\tfrac18\,\del_\mu\cH_{MN}\,\del^\mu\cH^{MN}\Big]    \;,
\end{split}
\end{equation}
 where $n$-dimensional spacetime indices are raised with the inverse metric $g^{\mu\nu}\,$. The abelian field strength is given by 
 \begin{equation}
\cF_{\mu\nu}{}^M=\del_\mu\cA_\nu{}^M-\del_\nu\cA_\mu{}^M   \;,  
 \end{equation}
while the three-form curvature $H_{\mu\nu\rho}$ needs an abelian Chern-Simons modification compared to the naive form $H=dB\,$:
\begin{equation}
H_{\mu\nu\rho}:=3\,\del_{[\mu}B_{\nu\rho]}-3\,\cA_{[\mu}{}^M\del_\nu\cA_{\rho]\,M}    \;,
\end{equation}
where the $O(d,d)$ indices have been contracted with the invariant metric $\eta_{MN}\,$.
The effective action \eqref{LEEA} is invariant under $n$-dimensional diffeomorphisms, as well as two-form gauge transformations $\delta_\zeta B_{\mu\nu}=\del_\mu\zeta_\nu-\del_\nu\zeta_\mu\,$. Invariance under the $U(1)^{2d}$ gauge transformations $\delta_\lambda\cA_\mu{}^M=\del_\mu\lambda^M$ requires the additional transformation of the $B$-field ${\delta_\lambda B_{\mu\nu}=\tfrac12\,\cF_{\mu\nu}{}^M\,\lambda_M}\,$.

The sigma model describing the coupling of the string to the spacetime fields is most easily written in terms of the $(n+d)$-dimensional field content as\footnote{For the moment we will ignore the coupling to the dilaton, since it is a higher order effect in $\alpha'\,$.}
\begin{equation}\label{Polyakov NOT constant}
S_{\rm string}=-\frac{1}{4\pi\alpha'}\int d^2\sigma\big[\sqrt{-h}h^{\alpha\beta}\,\hat G_{\hat\mu\hat\nu}(X)+\epsilon^{\alpha\beta}\,\hat B_{\hat\mu\hat\nu}(X)\big]\del_\alpha\hat X^{\hat\mu}\del_\beta\hat X^{\hat\nu}  \;.  
\end{equation}
The worldsheet fields split as $\hat X^{\hat\mu}=(X^\mu,Y^i)$, and it has been made explicit that the spacetime fields do not depend on $Y^i\,$. The $X^\mu$ coordinates obey periodic boundary conditions, $X^\mu(2\pi,\tau)=X^\mu(0,\tau)\,$, while the $Y^i(\sigma,\tau)$ in principle have winding contributions. However, in light of the discussion in the previous section, we shall restrict from the beginning to the sector with zero winding, \emph{i.e.} $Y^i(2\pi,\tau)=Y^i(0,\tau)\,$.

The $(n+d)$-dimensional fields are related to the $n$-dimensional ones by the usual Kaluza-Klein dictionary:
\begin{equation}\label{KKdictionaryG}
\begin{split}
&\hat G_{\mu\nu}=g_{\mu\nu}+A_\mu^i\,G_{ij}\,A_\nu^j\;,\quad G_{\mu i}=G_{ij}\,A_\mu^j\;,\quad \hat G_{ij}=G_{ij}  \;,\\
&\hat G^{\mu\nu}=g^{\mu\nu}\;,\quad \hat G^{\mu i}=-g^{\mu\nu}A_\nu^i\;,\quad \hat G^{ij}=G^{ij}+A_\mu^i\,g^{\mu\nu}A_\nu^j\;,
\end{split}    
\end{equation}
as well as
\begin{equation}\label{KKdictionaryB}
\begin{split}
&\hat B_{\mu\nu}=B_{\mu\nu}-A^i_{[\mu}\widetilde A_{\nu]\,i}+A_\mu^i\,B_{ij}\,A_\nu^j  \;,\\
&\hat B_{\mu i}=\widetilde A_{\mu\,i}-B_{ij}\,A_\mu^j\;,\quad \hat B_{ij}=B_{ij}\;.
\end{split}
\end{equation}
Using the reduction ansatz \eqref{KKdictionaryG} and \eqref{KKdictionaryB} directly in the Lagrangian action \eqref{Polyakov NOT constant} leads to a quite unintelligible mess. In the last section we have seen that the appearance of manifest  $O(d,d)$ invariance crucially relies on the Hamiltonian formalism. This suggests that the same should happen in the present context. We shall thus rewrite the action \eqref{Polyakov NOT constant} in Hamiltonian form. In terms of $(n+d)$-dimensional fields this does not require any different computation compared to \eqref{compactPs} and \eqref{Hamiltonaction}, thus giving for the momenta 
\begin{equation}\label{generalPs}
\hat P_{\hat\mu}=\frac{1}{2\pi \alpha'}\big[\tfrac{1}{e}\,\hat G_{\hat\mu\hat\nu}(\dot{ \hat X}^{\hat\nu}-u\,\hat X^{\hat\nu\,'})+\hat B_{\hat\mu\hat\nu}\,\hat X^{\hat\nu\,'}\big]\;,   \end{equation}
with $e$ and $u$ given as in \eqref{eu}, and
\begin{equation}\label{Hamiltonactionfull}
S_{\rm string}=\int d^2\sigma\,\Big[\hat P_{\hat\mu}\dot{ \hat X}^{\hat\mu}-e\,\cH-u\,\cN\Big]\;. 
\end{equation}
The Virasoro constraints $\cN$ and $\cH$ are also given by the same expressions as in \eqref{NH}, 
except that all quantities such as $G$ and $B$ are replaced  by hatted quantities  $\hat G$ and $\hat B\,$. 
Splitting the symplectic term and the $\cN$ constraint is trivial, since they do not contain spacetime fields:
\begin{equation}\label{OddN}
\hat P_{\hat\mu}\dot{ \hat X}^{\hat\mu}=P_\mu\dot X^\mu+P_i\dot Y^i\;,\quad \cN=P_\mu\del_\sigma X^\mu+P_i\del_\sigma Y^i=P_\mu\del_\sigma X^\mu+\tfrac{1}{4\pi\alpha'}\,Z^MZ_M\;,    
\end{equation}
where we recalled the vector $Z^M$ defined in (\ref{ZMvector}). 
The challenge is to express $\cH$  in terms of $n$-dimensional fields by using \eqref{KKdictionaryG} and \eqref{KKdictionaryB}. After a tedious computation the final result can be expressed in a manifest $O(d,d)$ invariant form:
\begin{equation}\label{OddH}
\begin{split}
\cH&=\frac{1}{4\pi\alpha'}\big\{g^{\mu\nu}\Pi_\mu\Pi_\nu-2\,g^{\mu\lambda}\cB_{\lambda\nu}\,\Pi_\mu\del_\sigma X^\nu+(g_{\mu\nu}+g^{\lambda\sigma}\cB_{\lambda\mu}\cB_{\sigma\nu})\del_\sigma X^\mu\del_\sigma X^\nu\\
&\hspace{5mm}+\cH_{MN}(Z^M+\cA_\mu{}^M \del_\sigma X^\mu)(Z^N+\cA_\nu{}^N \del_\sigma X^\nu)\big\}  \;,  
\end{split}    
\end{equation}
where
\begin{equation}
\Pi_\mu:=2\pi\alpha'\,P_\mu-\cA_\mu{}^M Z_M\;,\quad \cB_{\mu\nu}:=B_{\mu\nu}+\tfrac12\,\cA_\mu{}^M\cA_{\nu\,M} \;.   
\end{equation}
The Virasoro constraints $\cN$ and $\cH$ are clearly $O(d,d)$ invariant, modulo the issue of zero modes discussed in the previous section, that we will revisit in the present context.

There is no reason to keep the non-compact sector in Hamiltonian form. We shall thus eliminate the momenta $P_\mu$ by their equations of motion:
\begin{equation}\label{Pmuonshell}
2\pi\alpha'\,P_\mu=e^{-1}g_{\mu\nu}\,\mathring{X}^\nu+\cB_{\mu\nu}\,\del_\sigma X^\nu+\cA_\mu{}^M Z_M  \;,\qquad \mathring{X}^\mu:=\del_\tau X^\mu-u\,\del_\sigma X^\mu  \;,
\end{equation}
and recast the action \eqref{Hamiltonactionfull} in the mixed  form
\begin{equation}\label{Mixedactionfull}
\begin{split}
S_{\rm string}&=\frac{1}{2\pi\alpha'}\,\int d^2\sigma\,\Big[\tfrac{1}{2e}\,g_{\mu\nu}\,\mathring{X}^\mu \mathring{X}^\nu+\big(\cB_{\mu\nu}\del_\sigma X^\nu+\cA_\mu{}^M Z_M\big)\mathring{X}^\mu-\tfrac{e}{2}\,g_{\mu\nu}\,\del_\sigma X^\mu\del_\sigma X^\nu\\
&+2\pi\alpha'\,P_i\dot Y^i-\tfrac{u}{2}\,Z^MZ_M-\tfrac{e}{2}\,\cH_{MN}\,\big(Z^M+\cA_\mu{}^M \del_\sigma X^\mu\big)\big(Z^N+\cA_\nu{}^N \del_\sigma X^\nu\big)\Big]  \;.  
\end{split}    
\end{equation}

\subsection{Zero Mode Truncation}

As discussed in the previous section, the formal $O(d,d)$ invariance of the Virasoro constraints $\cH$ and $\cN$ is broken by the zero mode $p_{i\,0}=\frac{1}{2\pi}\int_0^{2\pi}d\sigma P_i$ even in the zero winding sector. Moreover, the symplectic term $p_{i\,0}\dot y^i_0$ is another $O(d,d)$ breaking term. Following the discussion in the previous section, we shall thus truncate the action\footnote{In order to ensure equivalence with the original action, one has to keep track of the $y^i_0$ equation of motion, which will be done in the following.} by projecting out the conjugate pair of zero modes. In order to do this, we set
\begin{equation}\label{Tseytlinfinal}
    2\pi\alpha'\,P_i=\del_\sigma\widetilde Y_i\;,
\end{equation}
that is an invertible field redefinition for the non-zero modes, and at the same time sets $p_{i\,0}=0$ upon taking $\widetilde Y_i(2\pi,\tau)=\widetilde Y_i(0,\tau)\,$. 
By using \eqref{Tseytlinfinal} one has $Z^M=\del_\sigma Y^M$ and the truncated symplectic current can be written in manifestly $O(d,d)$ invariant form:
\begin{equation}
\int d^2\sigma\,2\pi\alpha'\,P_i\dot Y^i=\tfrac12\int d^2\sigma \,\del_\sigma Y^M\del_\tau Y_M\;.   
\end{equation}

Before using \eqref{Tseytlinfinal} in \eqref{Mixedactionfull}, let us discuss spacetime gauge invariances. $n$-dimensional diffeomorphisms are a manifest invariance, upon using
\begin{equation}\label{n-diffeos}
\delta_\xi X^\mu=-\xi^\mu(X)\;,\quad \delta_\xi \Phi(X)=\cL_\xi\Phi(X)+\delta_\xi X^\mu\del_\mu\Phi(X) \;, 
\end{equation}
where $\Phi$ generically denotes spacetime fields, and we recalled that, when considering target space fields on the worldsheet, one has to add the extra term in \eqref{n-diffeos} to account for the explicit dependence on $X^\mu(\sigma, \tau)\,$. Invariance under two-form gauge transformations $\delta_\zeta B_{\mu\nu}=2\,\del_{[\mu}\zeta_{\nu]}$ is also standard. The situation is more subtle for the vector gauge symmetries $\delta_\lambda\cA_\mu{}^M=\del_\mu\lambda^M\,$. The upper component $\delta A_\mu^i=\del_\mu\lambda^i$ is a remnant of $(n+d)$-dimensional diffeomorphisms. This already fixes the transformation for the internal worldsheet coordinates: $\delta_\lambda Y^i=-\lambda^i(X)\,$. In order to preserve $O(d,d)\,$, one is led to demand $\delta_\lambda\widetilde Y_i=-\tilde\lambda_i\,$, so that 
\begin{equation}\label{deltaY}
\delta_\lambda Y^M=-\lambda^M(X)\;.  \end{equation}
In terms of internal momenta, the lower component gives $\delta_\lambda P_i=-\frac{1}{2\pi\alpha'}\del_\sigma\tilde\lambda_i\,$, that can also be derived by its on-shell expression \eqref{generalPs} for $\hat\mu=i\,$. This transformation preserves the solution space with $p_{i\,0}=0\,$, since $\delta_\lambda p_{i\,0}=0\,$.

Following \cite{Maharana:1992my} we  introduce the gauge-invariant derivative
\begin{equation}\label{DY}
D_\alpha Y^M:=\del_\alpha Y^M+\cA_\mu{}^M(X)\del_\alpha X^\mu   \;,
\end{equation}
which indeed obeys $\delta_\lambda(D_\alpha Y^M)=0$. 
The action \eqref{Mixedactionfull}, with truncated zero modes according to \eqref{Tseytlinfinal}, can be finally written as:\footnote{Recall that $e$ and $u$ are defined in 
terms of $h_{\alpha\beta}\,$ as  $e=\sqrt{-h}\,h_{\sigma\sigma}^{-1}$ and $u=h_{\tau\sigma}\,h_{\sigma\sigma}^{-1}\,$.}
\begin{equation}\label{finalaction}
\begin{split}
S=&-\frac{1}{4\pi\alpha'}\int d^2\sigma\,\Big[\sqrt{-h}\,h^{\alpha\beta}\,g_{\mu\nu}\,\del_\alpha X^\mu\del_\beta X^\nu+\epsilon^{\alpha\beta}\big(B_{\mu\nu} \del_\alpha X^\mu\del_\beta X^\nu-\cA_\mu{}^M D_\alpha Y_M\,\del_\beta X^\mu\big)\Big]  \\
&+\frac{1}{4\pi\alpha'}\,\int d^2\sigma\,\Big[D_\sigma Y^MD_\tau Y_M-u\,D_\sigma Y^MD_\sigma Y_M-e\,\cH_{MN}\,D_\sigma Y^MD_\sigma Y^N\Big]\;.
\end{split}
\end{equation}
Not only is $O(d,d)$ manifestly realized, but all terms in the action are gauge invariant under the vector symmetries. For the two terms involving the $B$-field and the 
bare vector $\cA_\mu{}^M\,$ one has to check that
\begin{equation}
\delta_\lambda\cA_\mu{}^M\,\epsilon^{\alpha\beta}D_\alpha Y_M\del_\beta X^\mu =\del_\alpha v^\alpha+\tfrac12\,\lambda_M\,\cF_{\mu\nu}{}^M\,\epsilon^{\alpha\beta}\del_\alpha X^\mu\del_\beta X^\nu \;,  
\end{equation}
which exactly cancels the non-standard transformation $\delta_\lambda B_{\mu\nu}=\tfrac12\,\cF_{\mu\nu}{}^M \lambda_M$ of the $B$-field, thus proving invariance of the action under the spacetime gauge symmetries.
The $O(d,d)$ symmetric action \eqref{finalaction} also has a manifest zero mode local symmetry under
\begin{equation}\label{Xisymmetry}
\delta_\Xi Y^M(\sigma,\tau)=\Xi^M(\tau)\;,    
\end{equation}
that will be used to show equivalence with the (truncated) original sigma model.

Having found the final form \eqref{finalaction} of the action, let us now show that it provides a consistent truncation of the original theory. 
To this end we have to determine  the zero mode $y_0^i$ in terms of the untruncated fields so that the original equations of motion are satisfied. 
The $i$-component of \eqref{generalPs} gives
\begin{equation}\label{compactPonshell}
2\pi\alpha'\,P_i=\tfrac{1}{e}\,G_{ij}\,(D_\tau Y^j-u\,D_\sigma Y^j)+B_{ij}\,D_\sigma Y^j-\tilde A_{\mu\,i}\,\del_\sigma X^\mu    
\end{equation}
upon using the Kaluza-Klein ansatz \eqref{KKdictionaryG}, \eqref{KKdictionaryB} and the definition \eqref{DY} of $D_\alpha Y^M\,$. As mentioned above, this also confirms the transformation law $\delta_\lambda P_i=-\tfrac{1}{2\pi\alpha'}\del_\sigma\tilde\lambda_i$ under the vector gauge symmetries. Inverting \eqref{compactPonshell} and integrating over $\sigma$ one obtains the original equation for the zero mode $y^i_0\,$:
\begin{equation}\label{y0equation}
\dot y^i_0(\tau)=V^i(\tau)\;,    
\end{equation}
with
\begin{equation}\label{Y0source}
V^M(\tau)=\frac{1}{2\pi}\int_0^{2\pi}d\sigma\,\Big[u\,D_\sigma Y^M+e\,\cH^{MN}D_\sigma Y_N-\cA_\mu{}^M\,\del_\tau X^\mu\Big]\;,  \end{equation}
where \eqref{Tseytlinfinal} has been used to ensure $p_{i\,0}=0\,$.
Integration of \eqref{y0equation} then determines $y^i_0(\tau)$ in terms of the untruncated fields consistent with the equations of motion. 
One may also verify that \eqref{y0equation} is invariant under the spacetime gauge symmetry. We note that only the last term above has a non-trivial transformation under $U(1)\,$, explicitly 
\begin{equation}
\delta_\lambda V^i=-\frac{1}{2\pi}\int_0^{2\pi}d\sigma\del_\mu\lambda^i\del_\tau X^\mu=-\frac{1}{2\pi}\int_0^{2\pi}d\sigma\dot\lambda^i\;. 
\end{equation}
This ensures gauge invariance of \eqref{y0equation}, given that $y_0^i=\frac{1}{2\pi}\int_0^{2\pi}d\sigma\,Y^i$ and $\delta_\lambda
 Y^i=-\lambda^i(X)\,$. Similarly to the simpler case discussed in the previous section, one could also fix the (arbitrary) function $\tilde y_{i\,0}$ by supplementing the action \eqref{finalaction} with the manifestly $O(d,d)$ and spacetime gauge invariant extra equation
 \begin{equation}\label{Y0equation}
\dot Y^M_0=V^M\;.   \end{equation}
The extra condition \eqref{Y0equation} 
can be viewed as a gauge fixing condition for the $\Xi^M$ symmetry \eqref{Xisymmetry}. In this respect, the solutions of \eqref{finalaction} 
can be embedded into solutions of the original sigma model, up to gauge equivalence.

We end this section by examining the field equations obtained from the action \eqref{finalaction}. 
The $Y$ field equations are given by a total $\sigma$-derivative:
\begin{equation}\label{dsigmaduality}
\del_\sigma\Big[D_\tau Y^M-u\,D_\sigma Y^M-e\,\cH^{MN}D_\sigma Y_N\Big]=0\;,
\end{equation}
which makes explicit that the action \eqref{finalaction} does not determine the dynamics of the zero modes $Y^M_0(\tau)\,$. According to \eqref{dsigmaduality}, the quantity in brackets can be an arbitrary function of $\tau\,$, say $C^M(\tau)\,$, depending on the $\Xi$-gauge. It is easy to see that the $C^M(\tau)$ corresponding to the gauge choice \eqref{Y0equation} is $C^M=0\,$, yielding
\begin{equation}\label{dualityarbitrarygauge}
D_\tau Y^M-u\,D_\sigma Y^M-e\,\cH^{MN}D_\sigma Y_N=0\;,     
\end{equation}
that in conformal gauge $(e,u)=(1,0)$ takes the form of a covariantized self-duality relation:
\begin{equation}\label{dualityconformalgauge}
D_\alpha Y^M=\epsilon_\alpha{}^\beta\,\cH^{MN}D_\beta Y_N\;.    
\end{equation}
The first order equation \eqref{dualityarbitrarygauge} is physically equivalent to the gauge invariant variational equation \eqref{dsigmaduality}. However, it should be kept in mind that it can be used only when discussing pure on-shell properties in a fixed $\Xi^M$ gauge, and not otherwise.

The field equations for $X^\mu$ resulting from the action \eqref{finalaction} read
\begin{equation}\label{Xequationsarbitrarygauge}
\begin{split}
&g_{\mu\nu}\,\big(\nabla^2 X^\nu+\Gamma^\nu_{\lambda\rho}\,\nabla^\alpha X^\lambda\nabla_\alpha X^\rho\big)-\tfrac12\,\varepsilon^{\alpha\beta}\,\big[\del_\alpha X^\nu\del_\beta X^\lambda\,H_{\mu\nu\lambda}+2\,\del_\alpha X^\nu\,D_\beta Y^M\,\cF_{\mu\nu\,M}\big]\\
&+\tfrac{1}{\sqrt{-h}}\,\del_\sigma X^\nu\,\cF_{\mu\nu}{}^M\,\Big[D_\tau Y_M-u\,D_\sigma Y_M-e\,\cH_{MN}\,D_\sigma Y^N\Big]-\tfrac{e}{2\sqrt{-h}}\,\del_\mu\cH_{MN}\,D_\sigma Y^MD_\sigma Y^N=0\;,
\end{split}    
\end{equation}
where $\varepsilon^{\alpha\beta}:=\frac{1}{\sqrt{-h}}\epsilon^{\alpha\beta}\,$, and $\nabla_\alpha$ denote worldsheet covariant derivatives built from $h_{\alpha\beta}\,$.
In conformal gauge, and using the $\Xi$-gauge yielding the first-order equation \eqref{dualityarbitrarygauge}, this reduces to a result of Maharana and Schwarz \cite{Maharana:1992my}:
\begin{equation}\label{Xequationconformalgauge}
\begin{split}
&g_{\mu\nu}\,\big(\Box X^\nu+\Gamma^\nu_{\lambda\rho}\,\del^\alpha X^\lambda\del_\alpha X^\rho\big)-\tfrac12\,\epsilon^{\alpha\beta}\,\big[\del_\alpha X^\nu\del_\beta X^\lambda\,H_{\mu\nu\lambda}+2\,\del_\alpha X^\nu\,D_\beta Y^M\,\cF_{\mu\nu\,M}\big]\\
&-\tfrac{1}{4}\,\del_\mu\cH_{MN}\,D_\alpha Y^MD^\alpha Y^N=0\;,
\end{split}    
\end{equation}
where we used again the first order duality relation \eqref{dualityconformalgauge} to recast the last term in a manifestly Lorentz invariant form.

Finally, the equations of motion of the worldsheet metric, obtained by varying with respect to $e$ and $u$, are given by 
\begin{equation}
\begin{split}
-\frac{\delta S}{\delta e}&=\cH=\frac{1}{4\pi\alpha'}\,\Big[e^{-2}\,g_{\mu\nu}\,\mathring{X}^\mu\mathring{X}^\nu+g_{\mu\nu}\,\del_\sigma X^\mu\del_\sigma X^\nu+\cH_{MN}\,D_\sigma Y^M D_\sigma Y^N\Big]\;,\\
-\frac{\delta S}{\delta u}&=\cN=\frac{1}{4\pi\alpha'}\,\Big[2\,e^{-1}\,g_{\mu\nu}\,\mathring{X}^\mu\del_\sigma X^\nu+D_\sigma Y^MD_\sigma Y_M\Big]\;,
\end{split}    
\end{equation}
where $\mathring{X}^\mu=\del_\tau X^\mu-u\,\del_\sigma X^\mu$. 

In the whole discussion so far we glossed over the fate of worldsheet diffeomorphisms. In an appendix  we examine this issue in great detail and provide the explicit realization of diffeomorphism symmetry in the action \eqref{finalaction}. In particular, we prove that the diffeomorphism transformations 
\begin{equation}
\begin{split}
\d_{\xi}X^{\m}&=\xi^{\a}\del_{\a}X^{\m},\\
\delta_{\xi} Y^M&=\xi^\alpha\del_\alpha Y^M-\xi^\tau\,\Big[D_\tau Y^M-u\,D_\sigma Y^M-e\,\cH^{MN}\,D_\sigma Y_N\Big]    \;, 
\end{split}
\end{equation}
are an off-shell invariance of the action.

\section{Anomalies}

In the previous sections we have constructed the manifestly $O(d,d)$ invariant worldsheet sigma model \eqref{finalaction}. Worldsheet diffeomorphism invariance is not manifest, but it is extensively discussed in an appendix, as is the cancellation of gravitational anomalies.

The sigma model \eqref{finalaction} seems a good starting point to perform worldsheet perturbation theory in an $O(d,d)$ covariant way, to all orders in $\alpha'\,$. However, from the analysis of the low-energy spacetime theory,  it has been recently found \cite{Eloy:2019hnl,Eloy:2020dko} that the $B-$field
acquires a non-trivial transformation under $O(d,d)$ at first order in $\alpha'\,$. This is reminiscent of the original Green-Schwarz mechanism \cite{Green:1984sg} in type I or heterotic string theory. Similarly to the heterotic worldsheet theory that contains chiral fermions in both the gravitational and gauge sectors, the $O(d,d)$ sigma model \eqref{finalaction} is defined in terms of chiral bosons \`{a} la Floreanini-Jackiw \cite{Floreanini:1987as}. This suggests that the novel $O(d,d)$ Green-Schwarz mechanism found in \cite{Eloy:2019hnl} can also be explained in terms of worldsheet anomalies, as we will show here.

\subsection{Frame-like Worldsheet Action}

In this section we will focus on the $Y$ sector of the sigma model. The aim is to exhibit  two-dimensional anomalies that underlie the $O(d,d)$ Green-Schwarz deformation. 
Since the $\alpha'$ deformation of \cite{Eloy:2019hnl} does not involve the Kaluza-Klein gauge fields $\cA_\mu{}^M\,$, we will set them to zero and focus on the action
\begin{equation}\label{SYnonlin}
S_Y=\frac{1}{4\pi\alpha'}\int d^2\sigma\,\Big[\del_\sigma Y^M\,\del_\tau Y_M-\cH_{MN}(X)\,\del_\sigma Y^M\del_\sigma Y^N\Big]\;.    
\end{equation}

It is convenient to rewrite this action in terms of a frame formalism, which we briefly introduce now. 
The generalized metric can be 
written in terms of frame fields as \cite{Siegel:1993th, Siegel:1993xq, Hohm:2010pp}
\begin{equation}
\label{hMNflat}
\cH_{MN}(x)={E_{M}}^{A}(x)\, h_{AB}\, {E_{N}}^{B}(x)\;,
\end{equation}
where we have introduced the frame field ${E_{M}}^{A}$, and a $SO(d)\times SO(d)$ constant metric $h_{AB}$. The $O(d,d)$ invariant metric $\y_{MN}$, on the other hand, can be written as
\begin{equation}
\label{OddonE}
\y_{MN}={E_{M}}^{A}(x)\, \y_{AB}\, {E_{N}}^{B}(x)\;,
\end{equation}
where $\eta_{AB}$ has the same numerical form as $\eta_{MN}\,$. This choice implies that the frame field itself is an $O(d,d)$ matrix.
In the following we will use $\y^{AB}$ and $\eta_{AB}$ to raise and lower flat indices. Denoting  the inverse vielbein by ${E_{A}}^{M}$, such that 
${E_{M}}^{A}{E_{A}}^{N}={\d_{M}}^{N}$ and ${E_{A}}^{M}{E_{M}}^{B}={\d_{A}}^{B}$, the raising and lowering of indices is then consistent 
with taking inverses: 
\begin{equation}\label{inverseErules}
E_A{}^M=\eta_{AB} \,\eta^{MN} E_N{}^B\;.    
\end{equation}
Furthermore, $h_{AB}$ satisfies the constraints $h_{AC}\eta^{CD}h_{CB}=\eta_{AB}$ and $\eta^{AB}h_{AB}=0$.
In this formalism one has in addition to rigid $O(d,d)$ 
transformations 
 local $SO(d)\times SO(d)$ transformations:
\begin{equation}\label{localsod}
\delta_\lambda E_M{}^A(x)=-\lambda_B{}^A(x)\,E_M{}^B(x)\;,\quad \delta_\lambda E_A{}^M(x)=\lambda_A{}^B(x)\,E_B{}^M(x)\;, 
\end{equation}
where  the parameters $\lambda_A{}^B$ obey the $SO(d)\times SO(d)$ condition $\lambda_{(A}{}^C\,h_{B)C}=0$ 
and the $O(d,d)$ condition $\lambda_{(A}{}^C\,\eta_{B)C}=\lambda_{(AB)}=0$.    
The $SO(d)\times SO(d)$ preserving condition on $\lambda$ can be conveniently rewritten as
\begin{equation}\label{lambdasod}
\lambda_{AC}\,h^C{}_B+\lambda_{BC}\,h^C{}_A=\lambda_{AC}\,h^C{}_B-h^C{}_A\,\lambda_{CB}=[\lambda,h]_{AB}=0\;,    
\end{equation}
where we used $h=h^{\rm T}$ as a matrix.

In the following it will be important to separate irreducible $SO(d)\times SO(d)$ representations from any tensor with indices $A,B=1,\ldots, 2d$. 
This can be achieved by use of projection operators  
\begin{equation}
\Pi_{\pm}^A{}_B=\frac12 \big(\delta^A{}_B\pm h^A{}_B\big)    \,,
\end{equation}
which, thanks to the constraints stated after (\ref{inverseErules}), 
are orthogonal and obey $\Pi_\pm^2=\Pi_\pm$, the completeness relation  $\mathbf{1}=\Pi_++\Pi_-$ and ${\rm Tr}\,\Pi_\pm=d$. 
This allows us to decompose an arbitrary vector $V^A$ as
\begin{equation}
V^A=V_+^A+V_-^A=V^{\un A}+V^{\ov A}\;, 
\end{equation}
where we shall denote by $\un A$ an index projected via $\Pi_+$ and $\ov A$ an index projected by $\Pi_-\,$. This way $\un A$ and $\ov A$ indices carry the $(d,0)$ and $(0,d)$ representations of $SO(d)\times SO(d)\,$, respectively.
Higher tensors decompose analogously. For instance, the gauge parameter $\lambda_{AB}$ decomposes as 
\begin{equation}
\lambda_{AB}=\lambda_{\un A\un B}+\lambda_{\ov A\ov B}\;,
\end{equation}
with $\lambda_{(\un A\un B)}=0, \lambda_{(\ov A\ov B)}=0$, 
where the vanishing of the off-diagonal components $\lambda_{\un A\ov B}$ and $\lambda_{\ov A\un B}$ follows since $[\lambda,\Pi_\pm]=0$ by  \eqref{lambdasod}. This fact makes it manifest that the gauge group is only $SO(d)\times SO(d)$.

Let us finally define the (composite) gauge fields for the $SO(d)\times SO(d)$ gauge symmetry. 
We start from the Maurer-Cartan form
\begin{equation}\label{WAB}
W_{\mu\,AB}:= E_A{}^M\del_\mu E_{MB}=-W_{\mu\,BA}\,,
\end{equation}
that can be decomposed into connections $Q_{\mu\,AB}$ of $SO(d)\times SO(d)\,$:
\begin{equation}
\begin{split}
&Q_{\mu\,AB}:=(\Pi_+W_\mu\Pi_+)_{AB}+(\Pi_-W_\mu\Pi_-)_{AB}=Q_{\mu\,\un A\un B}+Q_{\mu\,\ov A\ov B} \;,\\
\end{split}
\end{equation}
satisfying $Q_{\mu\,(\un A\un B)}=0$, $Q_{\mu\,(\ov A\ov B)}=0$,  
and a tensor $P_{\mu\,AB}$ in the $(d,d)$ representation:
\begin{equation}
\begin{split}
&P_{\mu\,AB}:=(\Pi_+W_\mu\Pi_-)_{AB}+(\Pi_-W_\mu\Pi_+)_{AB}=P_{\mu\,\un A\ov B}+P_{\mu\,\ov A\un B} \;,\\
\end{split}
\end{equation}
where $P_{\mu\,\ov A\un B}=-P_{\mu\,\un B\ov A}$. 
More precisely,  the transformation properties under \eqref{localsod} are:
\begin{equation}
\begin{split}
&\delta_\lambda Q_{\mu\,\un A\un B}=-\calD_\mu\lambda_{\un A\un B}:= -\Big(\del_\mu\lambda_{\un A\un B}+[Q_\mu,\lambda]_{\un A\un B}\Big)\;,\\    
&\delta_\lambda Q_{\mu\,\ov A\ov B}=-\calD_\mu\lambda_{\ov A\ov B}:= -\Big(\del_\mu\lambda_{\ov A\ov B}+[Q_\mu,\lambda]_{\ov A\ov B}\Big)\;,\\
&\delta_\lambda P_{\mu\,\un A\ov B}=\lambda_{\un A}{}^{\un C}\,P_{\mu\,\un C\ov B}+\lambda_{\ov B}{}^{\ov C}\,P_{\mu\,\un A\ov C}\;,
\end{split}
\end{equation}
or, without splitting, $\delta_\lambda Q_{\mu\,AB}=-\calD_\mu\lambda_{AB}$ and $\delta_\lambda P_{\mu\,AB}=[\lambda,P_{\mu}]_{AB}\,$. 
Finally, the Maurer-Cartan form obeys the zero curvature identity $dW+W^2=0$, which 
gives rise to the Bianchi identities 
\begin{equation}
\begin{split}
R_{\mu\nu}&\equiv  \del_\mu Q_{\nu}-\del_\nu Q_{\mu}+[Q_\mu, Q_\nu]=-[P_\mu,P_\nu]\;,\\    
\calD_{[\mu}P_{\nu]}&\equiv  \del_{[\mu}P_{\nu]}+[Q_{[\mu},P_{\nu]}]=0\;, 
\end{split}    
\end{equation}
where we used matrix notation.

After this review of the frame formalism we now return to  the worldsheet theory \eqref{SYnonlin}. We perform the field redefinition that 
flattens the worldsheet fields $Y^M$: 
\begin{equation}
Y^M=E_A{}^M(X) Y^A\;.    
\end{equation}
The worldsheet derivatives $\del_\alpha Y^M$ then become
\begin{equation}
\del_\alpha Y^M=E_A{}^M\,(\del_\alpha Y^A+W_{\alpha}{}^A{}_B\,Y^B)=E_A{}^M\hat\calD_\alpha Y^A  \;, 
\end{equation}
where we introduced the pullback $W_{\alpha\,AB}:=\del_\alpha X^\mu\,W_{\mu\,AB}$ and the hatted covariant derivative \begin{equation}
\hat\calD_\alpha Y^A:=\del_\alpha Y^A+W_\alpha{}^A{}_B\,Y^B=\calD_\alpha Y^A+P_\alpha{}^A{}_B\,Y^B  \;,  
\end{equation}
that differs from the $SO(d)\times SO(d)$ covariant derivative $\calD_\alpha$, which is defined by this equation,  by the above coupling to $P_{\alpha\,AB}\,$.
The action \eqref{SYnonlin} can thus be written as
\begin{equation}\label{SYDhats}
S_Y=\frac{1}{4\pi\alpha'}\int d^2\sigma\,\Big[\hat\calD_\sigma Y^A\,\hat\calD_\tau Y_A-h_{AB}\,\hat\calD_\sigma Y^A\,\hat\calD_\sigma Y^B\Big] \;.   
\end{equation}
The zero-mode symmetry $\delta_\Xi Y^M=\Xi^M\,$, with $\del_\sigma\Xi^M=0\,$, now turns into
\begin{equation}
\delta_\Xi Y^A=\Xi^A\;,\quad \text{where} \quad \hat\calD_\sigma\Xi^A=0\;.    
\end{equation}
By means of the projectors $\Pi_\pm$ one can split $Y^A$ into $SO(d)\times SO(d)$ representations: $Y^A=Y^{\un A}+Y^{\ov A}$ under which the action decomposes as
\begin{equation}\label{SYdecomposed}
\begin{split}
S_Y&=\frac{1}{2\pi\alpha'}\int d^2\sigma \,\Big[\hat\calD_\sigma Y^{\un A}\,\hat\calD_-Y_{\un A}+\hat\calD_\sigma Y^{\ov A}\,\hat\calD_+Y_{\ov A}\Big]\\[2mm]
&=\frac{1}{2\pi\alpha'}\int d^2\sigma \,\Big[\big(\calD_\sigma Y^{\un A}+P_\sigma{}^{\un A}{}_{\ov B}\,Y^{\ov B}\big)\big(\calD_- Y_{\un A}+P_{-\,\un A\ov C}\,Y^{\ov C}\big)\\
&\hspace{22mm}+\big(\calD_\sigma Y^{\ov A}+P_\sigma{}^{\ov A}{}_{\un B}\,Y^{\un B}\big)\big(\calD_+ Y_{\ov A}+P_{+\,\ov A\un C}\,Y^{\un C}\big)\Big] \;, 
\end{split}
\end{equation}
with $\calD_\pm=\frac12(\calD_\tau\pm\calD_\sigma)\,$.
Under a local $SO(d)\times SO(d)$ transformation, 
\begin{equation}
\delta_\lambda Y^A=\lambda^A{}_B(X)\,Y^B\;,    
\end{equation}
the hatted derivatives $\hat\calD_\alpha Y^A$ transform covariantly, \emph{i.e.}
\begin{equation}
\delta_\lambda(\hat\calD_\alpha Y^A)=\lambda^A{}_B\,\hat\calD_\alpha Y^B \;. \end{equation}
It is thus clear that the action \eqref{SYDhats} is invariant under (spacetime) local $SO(d)\times SO(d)$, provided  one transforms simultaneously $Y^A$ and $W_\alpha^{AB}\,$.

We now turn to a general discussion of potential anomalies in this model, which will be computed explicitly in the next subsection. 
It must first be emphasized that the above invariance under $SO(d)\times SO(d)$ is not a genuine symmetry of the worldsheet theory, 
since the background fields (target space fields) need to be transformed as well. 
Nonetheless, it is an important consistency condition that any two configurations of target space fields that are gauge equivalent (from 
the target space point of view) give rise to equivalent worldsheet theories. It is this property that may become anomalous. 
This is precisely analogous to heterotic string theory where the coupling to target space Yang-Mills gauge fields is quantum-mechanically 
inconsistent unless the Green-Schwarz mechanism is invoked \cite{Hull:1985jv}. 

While the $SO(d)\times SO(d)$ is not a genuine symmetry of the worldsheet theory, one can derive consequences 
from this invariance property: 
\begin{equation}\label{conservationtrick}
\int d^2\sigma\,\Big[\frac{\delta S}{\delta W_\alpha^{AB}}\,\hat\calD_\alpha\lambda^{AB}\Big]=-\frac{1}{4\pi\alpha'}\int d^2\sigma\,\Big[\lambda^{AB}\,\hat\calD_\alpha\,\cJ^\alpha_{AB}\Big]=0\;, 
\end{equation}
where we assumed that the $Y$'s are on-shell: $\frac{\delta S}{\delta Y^A}=0\,$, and we have defined
\begin{equation}
\begin{split}
&\cJ^\alpha_{AB}:=4\pi\alpha'\,\frac{\delta S}{\delta W_\alpha^{AB}}\;,
\end{split}
\end{equation}
which reads in components 
\begin{equation}
\begin{split}
&\cJ^\tau_{AB}=-Y_{[A}\hat\calD_\sigma Y_{B]}\;,\quad \cJ^\sigma_{AB}=-Y_{[A}\big(\hat\calD_\tau Y_{B]}-2\,h_{B]C}\,\hat\calD_\sigma Y^C\big)\;.
\end{split}    
\end{equation}
While the $\cJ^\alpha_{AB}$ are not conserved  $SO(d)\times SO(d)$ currents 
the relation \eqref{conservationtrick} 
implies the projected `conservation' law 
\begin{equation}
\Pi_+^{AC}\Pi_+^{BD}\,(\hat\calD_\alpha\cJ^\alpha)_{CD}=0 \;,\quad \Pi_-^{AC}\Pi_-^{BD}\,(\hat\calD_\alpha\cJ^\alpha)_{CD}=0\;,  
\end{equation}
or, using manifest $SO(d)\times SO(d)$ indices, 
\begin{equation}\label{almostconserved}
\calD_\alpha\cJ^\alpha_{\un A\un B}+P_{\alpha\,\un A}{}^{\ov C}\cJ^\alpha_{\ov C \un B}-P_{\alpha\,\un B}{}^{\ov C}\cJ^\alpha_{\ov C \un A}=0 \;,\quad
\calD_\alpha\cJ^\alpha_{\ov A\ov B}+P_{\alpha\,\ov A}{}^{\un C}\cJ^\alpha_{\un C \ov B}-P_{\alpha\,\ov B}{}^{\un C}\cJ^\alpha_{\un C \ov A}=0\;.
\end{equation}
One can explicitly verify that \eqref{almostconserved} holds upon using the equation of motion
\begin{equation}
\hat\calD_\sigma\big(\hat\calD_\tau Y^A-h^A{}_B\,\hat\calD_\sigma Y^B\big)=0\;. \end{equation}
Despite \eqref{almostconserved} not being a standard conservation law,
the above result shows that the free theory (where $W_\mu^{AB}$ is set to zero) does have conserved $SO(d)\times SO(d)$ currents $j^\alpha_{\un A \un B}$ and $j^\alpha_{\ov A \ov B}\,$:
\begin{equation}\label{sodfree}
\begin{split}
j^\tau_{\un A\un B}=-Y_{[\un A}\del_\sigma Y_{\un B]} \;,\quad j^\sigma_{\un A\un B}=-Y_{[\un A}\big(\del_\tau-2\,\del_\sigma\big)Y_{\un B]} \;,\\  
j^\tau_{\ov A\ov B}=-Y_{[\ov A}\del_\sigma Y_{\ov B]} \;,\quad j^\sigma_{\ov A\ov B}=-Y_{[\ov A}\big(\del_\tau+2\,\del_\sigma\big)Y_{\ov B]}\;,
\end{split}    
\end{equation}
obeying the usual conservation law $\del_\alpha j^\alpha_{\un A\un B}=0\,$, $\del_\alpha j^\alpha_{\ov A\ov B}=0\,$. 
This emergence of conserved currents can be understood most clearly in the original form of the action (\ref{SYnonlin}). 
In the free limit ${\cal H}_{MN}$ reduces to a constant (its background value), which is invariant under a global $SO(d)\times SO(d)$, 
hence giving rise to conserved Noether currents, given by (\ref{sodfree}).

In order to employ the above action for a perturbative quantum field theory treatment let us inspect 
the linearized coupling to $W_\mu^{AB}$ 
\begin{equation}\label{Yperturbative}
\begin{split}
S_Y&=\frac{1}{2\pi\alpha'}\int d^2\sigma\Big[\del_\sigma Y^{\un A}\,\del_-Y_{\un A}+\del_\sigma Y^{\ov A}\,\del_+Y_{\ov A}+\tfrac12\,W_\alpha^{A B}  J^\alpha_{ A B}\Big]+{\cal O}(W^2)\\
&=\frac{1}{2\pi\alpha'}\int d^2\sigma\Big[\del_\sigma Y^{\un A}\,\del_-Y_{\un A}+\del_\sigma Y^{\ov A}\,\del_+Y_{\ov A}+\tfrac12\,Q_\alpha^{\un A\un B} j^\alpha_{\un A\un B}+\tfrac12\,Q_\alpha^{\ov A\ov B} j^\alpha_{\ov A\ov B}+\tfrac12\,P_\alpha^{\un A \ov B} t^\alpha_{\un A\ov B}\Big]+{\cal O}(W^2)\,,
\end{split}
\end{equation}
which involves the usual interaction term of gauge field and current, but also a coupling to $P_\mu^{AB}\,$, through  the $(d,d)$ tensor $t^\alpha_{\un A\ov B}$ defined as
\begin{equation}
t^\tau_{\un A\ov B}=Y^{\ov B}\,\del_\sigma Y^{\un A}-Y^{\un A}\,\del_\sigma Y^{\ov B} \;,\qquad t^\sigma_{\un A\ov B}=Y^{\ov B}\,\big(\del_\tau-2\,\del_\sigma \big)Y^{\un A}-Y^{\un A}\,\big(\del_\tau+2\,\del_\sigma\big) Y^{\ov B}   \;.
\end{equation}
In a perturbative treatment of the above action, one splits the `external' coordinate fields $X^\mu(\tau, \sigma)$ into a background $X^\mu_0$ plus fluctuations $\pi^\mu$ while, for the present purpose, the $Y^A$ can be treated as purely quantum. Among others, the action \eqref{Yperturbative} produces the worldsheet vertices $W_\mu^{AB}(X_0)\,\del_\alpha X^\mu_0\,J^\alpha_{AB}(Y)$ and $\del_\mu W_\nu^{AB}(X_0)\,\pi^\mu\del_\alpha\pi^\nu\,J^\alpha_{AB}(Y)\,$. In principle, these two can combine with the vertex $H_{\mu\nu\lambda}(X_0)\,\epsilon^{\alpha\beta}\del_\alpha X^\mu_0\,\del_\beta\pi^\nu\,\pi^\lambda\,$, arising  from the expansion of $\int_\Sigma B$ in \eqref{finalaction}, to produce a two-loop contribution of the schematic form
\begin{equation}\label{schematicalpha'}
\alpha'\,\del_\alpha X^\mu_0\del^\alpha X^\nu_0\,\Big[H_\mu{}^{\lambda\sigma}\,{\rm tr}(W\!\wedge dW)_{\nu\lambda\sigma}\Big] \;.   
\end{equation}
Its divergent part contributes to the $\beta-$functional of the metric $g_{\mu\nu}\,$, thus modifying the Einstein equation $R_{\mu\nu}=\frac14\,H_\mu{}^{\lambda\sigma}H_{\nu\lambda\sigma}+\cdots$ by the above $\alpha'$ correction.
This is analogous to the correction underpinning the original Green-Schwarz deformation \cite{Callan:1989nz} that results in the redefinition $H\to H+\frac{\alpha'}{4\pi}\,(\omega_3(A)-\omega_3(\omega))$. In the present case, the structure  \eqref{schematicalpha'} matches with the $\alpha'$ deformation found in \cite{Eloy:2019hnl}.

Let us now discuss the appearance of potential anomalies in a little more detail. 
Focussing on the $Q-$dependent part of \eqref{schematicalpha'}, the very existence and the precise structure of the correction is determined by the one-loop two-point functions
\begin{equation}
\left\langle j_\alpha^{\un A\un B}\,j_\beta^{\un C\un D}\right\rangle \;,\quad \left\langle j_\alpha^{\ov A\ov B}\,j_\beta^{\ov C\ov D}\right\rangle\;. \end{equation}
In the standard heterotic string context, the Green-Schwarz  deformation is driven by the worldsheet chiral anomaly: the components $J_+^{ab}$ of the gauge current and $J_-^{\mu\nu}$ of the Lorentz current vanish identically in the classical theory, due to the chiral nature of the fermions $\lambda^a$ and $\psi^\mu\,$. The chiral anomaly, however, gives rise to non-vanishing two-point functions $\langle J_+J_-\rangle$ at one-loop that are ultimately responsible for the Green-Schwarz  deformation.

The situation for the model \eqref{Yperturbative} is not exactly the same, but quite similar. Indeed, none of the currents \eqref{sodfree} vanish \emph{identically}, but half of them are classically trivial: when written in light-cone coordinates, one has
\begin{equation}
\begin{split}
&j_+^{\un A \un B}=Y^{[\un A}\big(\del_+-2\,\del_-\big)Y^{\un B]}\;,\qquad j_-^{\un A \un B}=Y^{[\un A}\del_-Y^{\un B]}\;,\\     
&j_+^{\ov A \ov B}=-Y^{[\ov A}\del_+Y^{\ov B]}\;,\qquad \qquad \;\;\;\; j_-^{\ov A \ov B}=Y^{[\ov A}\big(2\,\del_+-\del_-\big)Y^{\ov B]}\;.
\end{split}    
\end{equation}
Let us recall that the free-field equations $\del_\sigma\del_-Y^{\un A}=0\,$, $\del_\sigma\del_+Y^{\ov A}=0$ imply the chirality conditions $\del_-Y^{\un A}=0$ and $\del_+Y^{\ov A}=0$ except for the zero-modes $Y^A_0(\tau)\,$. These, however, are pure gauge, thanks to the free-field symmetry
$\delta_\Xi Y^A=\Xi^A(\tau)\,$, and can be fixed to zero. This shows that, upon gauge fixing, the classical currents obey
\begin{equation}\label{classicalWard}
\begin{split}
j_-^{\un A\un B}=0\;,\qquad\del_-j_+^{\un A\un B}=0\;,\\
j_+^{\ov A\ov B}=0\;,\qquad\del_+j_-^{\ov A\ov B}=0\;.
\end{split}
\end{equation}
Focussing on the left-moving sector, at the quantum level the two-point function $\langle j_+^{\un A\un B}\,j_+^{\un C\un D}\rangle$ is certainly non-vanishing, which implies  that the classical relations \eqref{classicalWard} cannot hold. 
Indeed, the naive Ward identities read
\begin{equation}\label{non-anomalousWard}
\begin{split}
&p_-\langle j_+^{\un A\un B}(p)\,j_+^{\un C\un D}(-p)\rangle+p_+\langle j_-^{\un A\un B}(p)\,j_+^{\un C\un D}(-p)\rangle=0\;,\\
&p_+\langle j_-^{\un A\un B}(p)\,j_-^{\un C\un D}(-p)\rangle+p_-\langle j_+^{\un A\un B}(p)\,j_-^{\un C\un D}(-p)\rangle=0\;,
\end{split}
\end{equation}
so that, if $\langle j_-\, j_+\rangle=0$ continues to hold, they  cannot be satisfied, leading to an anomaly.
In the following we will provide a 
scheme that ensures $j_-=0$ in all two-point functions,  
so that the above shows that $\partial_{-}j_+=0$ cannot be satisfied if $\langle j_+\, j_+\rangle$ is non-zero.

\subsection{Anomalies of  Floreanini-Jackiw}

We will now confirm the existence of an anomaly by computing the above two-point functions. 
It is sufficient to focus on the free part of the action, which consists of $d$ left-moving and $d$ right-moving 
so-called Floreanini-Jackiw bosons. Focusing first on the left-moving sector we 
consider the action
\begin{equation}\label{ONFJ}
S=\frac{1}{2\pi\alpha'}\int d^2x\,\del_-\phi^a\,\del_\sigma\phi_a    \,,
\end{equation}
where  $a=1,\ldots ,d$. Here we have changed notation to emphasize that the following holds generally for the Floreanini-Jackiw model. 
The action  is invariant under rigid $SO(d)$ transformations given by
\begin{equation}
\delta\phi^a=\lambda^{ab} \phi_b\;, \end{equation}
where indices are lowered and raised  with the $SO(d)$ metric $\delta_{ab}$ and its inverse, respectively. The Noether currents associated with $SO(d)$ are given by
\begin{equation}\label{son}
j_-^{ab}=\frac{1}{\alpha'}\,\phi^{[a}\del_-\phi^{b]}  \;,\qquad j_+^{ab}=\frac{1}{\alpha'}\,\phi^{[a}(\del_+-2\,\del_-)\phi^{b]}\,, \end{equation}
and are conserved, obeying $\partial_{+}j_{-}^{ab}+\partial_{-}j_{+}^{ab}=0$, thanks to the field equation
\begin{equation}
\del_\sigma\del_-\phi^a=(\del_+-\del_-)\del_-\phi^a=0\;.    
\end{equation}
Here and in what follows we shall denote the light-cone coordinates by $x^\pm=\tau\pm\sigma$, so that  $\del_\pm=\frac12\,(\del_\tau\pm\del_\sigma)\,$.  

The action and the field equations are Lorentz-invariant, with $\phi^a$ transforming in a non-standard way:\footnote{The action actually has the much larger symmetry 
$\delta \phi^a=\xi^{1}(x^+)\partial_1\phi^a+\xi^{-}(\tau)\partial_{-}\phi^a$, where $\xi^{1}$ and $\xi^{-}$ are arbitrary functions of their arguments. This is a manifestation of 
the infinite-dimensional conformal symmetry in two dimensions. Note that the `second Lorentz symmetry' with $\xi^{-}(\tau)=\lambda \tau$ is trivial in  the `chiral gauge'  
$\partial_{-}\phi^a=0$ that we shall employ and hence this symmetry will not play any role in what follows.}
\begin{equation}\label{nonstandardLorentz}
\delta_L\phi^a=\omega\,x^+\,(\del_+-\del_-)\phi^a=\omega\,x^+\,\del_\sigma\phi^a\;,
\end{equation}
in contrast with the scalar transformation
\begin{equation}
\delta_L\varphi=\omega\,(x^+\del_+-x^-\del_-)\varphi\;.    
\end{equation}
Under a Lorentz transformation, a standard one-form $A_\alpha$ transforms as
\begin{equation}
\delta_LA_\pm=\omega\,(x^+\del_+-x^-\del_-)A_\pm\pm\omega\,A_\pm  \;, 
\end{equation}
while the above current transforms according to
\begin{equation}\label{LorentzJtrans}
\delta_Lj_-^{ab}=\omega\,x^+(\del_+-\del_-)j_-^{ab}\;,\quad\delta_Lj_+^{ab}=\omega\,x^+(\del_+-\del_-)j_+^{ab}+\omega\,j_+^{ab}+\omega\,j_-^{ab}\;.    
\end{equation}

To be more precise, the above Lorentz invariance is present if the theory is defined on the plane, i.e., on two-dimensional Minkowski space, 
but we should recall that for the string we defined the theory on the cylinder, where  $\tau\in\mathbb{R}$ and $\sigma\in[0,2\pi]$.  
The fields can then be expanded in Fourier modes:
\begin{equation}\label{cylmodes}
\phi^a(\tau,\sigma)=\sum_{n\in\mathbb{Z}}\phi^a_n(\tau)\,e^{in\sigma}\;,\quad (\phi^a_n)^*=\phi^a_{-n}\,. 
\end{equation}
Since the Lorentz transformations (\ref{nonstandardLorentz}) depend explicitly on $\sigma$ they do not respect the periodicity conditions, 
and so there are no well-defined Lorentz transformations for the Fourier modes. 
Thus, on the cylinder Lorentz invariance is lost. 
It is also important to note that  the action \eqref{ONFJ} does not contain the zero-mode $\phi^a_0(\tau)\,$. This is reflected in the $\tau-$local symmetry
\begin{equation}
\delta_\xi\phi^a(\tau, \sigma)=\xi^a(\tau)\;,    
\end{equation}
that shifts $\phi_0^a$ by an arbitrary function, while leaving the non-zero modes $\phi_n^a$ inert.

Upon gauge fixing $\phi_0^a(\tau)=0\,$, the field equation is equivalent to the chirality  condition $\del_-\phi^a=0\,$, showing that one can consider the on-shell equivalent current
\begin{equation}
j_-^{ab}=0\;,\quad j_+^{ab}=\frac{1}{\alpha'}\,\phi^{[a}\del_+\phi^{b]}\;.
\end{equation}
Notice that the above current $j_\alpha^{ab}$ transforms as a chiral one-form, \emph{i.e.}
\begin{equation}
\delta_Lj_-^{ab}=0\;,\quad \delta_Lj_+^{ab}=\omega\,x^+\del_+j_+^{ab}+\omega\,j_+^{ab}    \,,
\end{equation}
only on-shell, upon using $\del_-\phi^a=0\,$.

By using the mode expansion \eqref{cylmodes} one can rewrite the free action as 
\begin{equation}\label{actionforquantummodes}
S=\frac{1}{2\alpha'}\int d\tau\,\sum_{n\neq0}\Big[in\,\phi_{a\,n}\,\dot\phi^a_{-n}-n^2\,\phi_{a\, n}\,\phi^a_{-n}\Big]  \;. 
\end{equation}
Upon decomposing the $\phi^a_{n}$ into two sets (i.e.~upon picking a polarization) this action can be brought immediately into Hamiltonian form, 
with the first term taking the standard $p\dot{q}$ form. 
It is then straightforward to perform canonical quantization, leading 
to the equal-time commutation relations
\begin{equation}
[\phi^a_n,\phi^b_m]=\frac{\alpha'}{m}\,\delta^{ab}\,\delta_{n+m}\;, \quad n,m\neq0 \;, \end{equation}
with the usual notation $\delta_{n+m}:=\delta_{n+m,0}\,$.
We assume from now on that the zero-mode has been gauge fixed to zero.
Ordinary creation-annihilation operators are defined as
\begin{equation}
A_n^a:=\sqrt{\frac{n}{\alpha'}}\,\phi^a_{-n}\;,\quad A_n^{\dagger\,a}:=\sqrt{\frac{n}{\alpha'}}\,\phi^a_{n}\;,\quad n>0    \,,
\end{equation}
and obey
\begin{equation}
[A_n^a, A_m^{\dagger\,b}]=\delta^{ab}\,\delta_{nm}  \;.  
\end{equation}
The mode expansion of the quantum fields $\phi^a$ can thus be written as
\begin{equation}
\phi^a(\tau,\sigma)=\sum_{n=1}^\infty\sqrt{\frac{\alpha'}{n}}\,\Big(A_n^{\dagger\,a}(\tau)\,e^{in\sigma}+A_n^{a}(\tau)\,e^{-in\sigma}\Big)\;.    
\end{equation}
This allows us to compute the equal-time commutator for the $\phi^a$ fields:
\begin{equation}\label{NLcommutator}
[\phi^a(\tau,\sigma_1), \phi^b(\tau,\sigma_2)]=-2\pi\alpha'i\,\delta^{ab}\,\epsilon(\sigma_1-\sigma_2)   \;, 
\end{equation}
with 
\begin{equation}
\epsilon(x):=\frac{1}{2\pi i}\sum_{n\neq0}\frac1n\,e^{inx}  \;,  
\end{equation}
obeying
\begin{equation}
\epsilon'(x)=\delta(x)-\frac{1}{2\pi}\;,\quad\epsilon(x)=-\epsilon(-x)\;.   \end{equation}
This clearly shows that the theory is non-local in $\sigma\,$, given that two fields at separated points do not commute at equal times. 

The quantum Hamiltonian can be read off from \eqref{actionforquantummodes} as
\begin{equation}
H=\frac{1}{2\alpha'}\,\sum_{n\neq0}n^2 :\phi_{a\, n}\,\phi^a_{-n}:    
\end{equation}
and allows one to compute the Heisenberg equation:
\begin{equation}
\dot\phi^a_n=i\,[H, \phi^a_n]=in\,\phi^a_n\;.  \end{equation}
This leads to the on-shell expansion
\begin{equation}\label{onshellexp}
\phi^a(\tau,\sigma)=\sum_{n=1}^\infty\sqrt{\frac{\alpha'}{n}}\,\Big(A_n^{\dagger\,a}\,e^{inx^+}+A_n^{a}\,e^{-inx^+}\Big)\;,  
\end{equation}
showing that the spectrum contains purely left-moving massless excitations.
With the above expansion one can compute the Feynman propagator:
\begin{equation}
\begin{split}
\Delta^{ab}(\tau,\sigma)&=\langle0|\,T\big\{\phi^a(\tau,\sigma)\,\phi^b(0,0)\big\}\,|0\rangle=\alpha'\,\delta^{ab}\,\Delta(\tau,\sigma)\,, \\
\Delta(\tau,\sigma)&=\Big(\theta(\tau)\,\sum_{n=1}^\infty\frac1n\,e^{-inx^+}+\theta(-\tau)\,\sum_{n=1}^\infty\frac1n\,e^{inx^+}\Big)\;,
\end{split}  
\end{equation}
that can be represented as
\begin{equation}\label{Feynman}
\Delta(x)=\frac{i}{4\pi}\int d^2k\,\frac{e^{ik\cdot x}}{k_-k_1+i\epsilon}  \;,  
\end{equation}
where we denoted $x^\alpha=(\tau,\sigma)\,$, $k_\alpha=(\omega,n)$ and
\begin{equation}
\int d^2k:=\int_{-\infty}^{+\infty}d\omega\,\sum_{n\neq 0}  \;. \end{equation}

The currents \eqref{son} do not suffer from ordering ambiguities at the quantum level, due to antisymmetrization in the $SO(d)$ indices. We can thus consider $j_-^{ab}=0$ and $j_+^{ab}=\frac{1}{\alpha'}\,\phi^{[a}\del_+\phi^{b]}$ as quantum operators and use the on-shell expansion \eqref{onshellexp} to compute the only non-vanishing two-point function
\begin{equation}
G_{++}^{ab,cd}(x-y):=\langle0|T\{j_+^{ab}(x)j_+^{cd}(y)\}|0\rangle  \;. \end{equation}
Writing the time-ordered product explicitly we have
\begin{equation}
G_{++}^{ab,cd}=\theta(\tau_1-\tau_2)\,\langle0|j_+^{ab}(x)j_+^{cd}(y)|0\rangle+\theta(\tau_2-\tau_1)\,\langle0|j_+^{cd}(y)j_+^{ab}(x)|0\rangle \;,   
\end{equation}
where $x^\alpha=(\tau_1,\sigma_1)$ and $y^\alpha=(\tau_2,\sigma_2)\,$. We then focus on the first factor and use the mode expansion to obtain
\begin{equation}
\begin{split}
\langle0|j_+^{ab}(x)j_+^{cd}(y)|0\rangle&=\sum_{n,m=1}^\infty\sum_{p,q=1}^\infty\sqrt{\frac{m\,q}{n\,p}}\langle0|A_n^{[a}A_m^{b]}\,e^{-i(n+m)x^+}A_p^{\dagger[c}A_q^{\dagger d]}\,e^{i(p+q)y^+}|0\rangle \\
&=\sum_{n,m=1}^\infty\sum_{p,q=1}^\infty\sqrt{\frac{m\,q}{n\,p}}e^{-i(n+m)x^++i(p+q)y^+}\delta^{c[a}\delta^{b]d}(\delta_{np}\delta_{mq}-\delta_{mp}\delta_{nq})\\
&=\tfrac12\,(\delta^{ac}\delta^{bd}-\delta^{bc}\delta^{ad})\sum_{n,m=1}^\infty\left(\frac{m}{n}-1\right)e^{-i(n+m)(x^+-y^+)}\\
&=\tfrac12\,(\delta^{ac}\delta^{bd}-\delta^{bc}\delta^{ad})\,F(x^+-y^+)\;,
\end{split}    
\end{equation}
where we defined 
\begin{equation}\label{F}
F(x):=\sum_{n,m=1}^\infty\left(\frac{m}{n}-1\right)e^{-i(n+m)x}  \,.   
\end{equation}
In order to evaluate this function
one can perform the sums on $n$ and $m$ separately. By using 
\begin{equation}
\begin{split}
&\sum_{n=1}^\infty z^n=\frac{z}{1-z} \;,\quad\sum_{n=1}^\infty n\, z^n=z\,\frac{d}{dz}\frac{1}{1-z}=\frac{z}{(1-z)^2} \;,  \\
&\sum_{n=1}^\infty\frac1n\,z^n=g(z)\;,\quad \frac{dg}{dz}=\sum_{n=1}^\infty z^{n-1}=\frac{1}{1-z}\;\rightarrow\;g(z)=-\log(1-z)\,,
\end{split}    
\end{equation}
one obtains
\begin{equation}
F(x)=-\frac{z}{(1-z)^2}\,\Big(z+\log(1-z)\Big)   \;,\quad z=e^{-ix} \,,
\end{equation}
or
\begin{equation}
F(x)=-\frac{1}{(e^{ix}-1)^2}\Big[1+e^{ix}\,\log(1-e^{-ix})\Big]\;.    
\end{equation}
In order to write the two-point function in momentum space, it is more useful to rewrite the double sum in \eqref{F} as
\begin{equation}\label{rewriting}
\sum_{n,m=1}^\infty f(n,m)=\sum_{N=2}^\infty\sum_{n=1}^{N-1} f(n, N-n) \;,   
\end{equation}
with $N=n+m\,$, yielding\footnote{The sum over $N$ can be extended to $N=1$ since $f(1)=0\,$.}
\begin{equation}\label{Fexp}
F(x)=\sum_{N=1}^\infty f(N)\,e^{-iNx}\;,    
\end{equation}
with
\begin{equation}\label{fresult}
f(N)=\sum_{n=1}^{N-1}\left(\frac{N}{n}-2\right)=N\,\Big[\psi(N)+\gamma\Big]-2(N-1)  \;,  
\end{equation}
where $\psi(x)=\frac{\Gamma'(x)}{\Gamma(x)}$ is the digamma function and $\gamma$ the Euler-Mascheroni constant.
It should be mentioned that the rewriting \eqref{rewriting} is not ambiguous, in that both forms of the series converge to the same function $F(x)\,$, that is regular for $x\neq0\,$.

The full two-point function can thus be written as
\begin{equation}\label{G++}
G_{++}^{ab,cd}(x)=\tfrac12\,(\delta^{ac}\delta^{bd}-\delta^{bc}\delta^{ad})\,\Big[\theta(\tau)\,F(x^+)+\theta(-\tau)\,F(-x^+)\Big]=\tfrac12\,(\delta^{ac}\delta^{bd}-\delta^{bc}\delta^{ad})\,G_{++}(x) \;. \end{equation}
At this point one can use \eqref{Fexp}, together with the integral representation for the step function:
\begin{equation}
\theta(\pm\tau)=\pm\frac{i}{2\pi}\int d\omega\,\frac{e^{-i\omega\tau}}{\omega\pm i\epsilon}\;,    
\end{equation}
in order to obtain the correlator in momentum space:
\begin{equation}\label{anomalouscorrelator}
G_{++}(p):=\int\frac{d^2x}{(2\pi)^2}\,e^{-ip_\alpha x^\alpha}G_{++}(x)=\frac{i}{4\pi p_-}\,\Big[p_1\,\big(\psi(|p_1|)+\gamma\big)-2\,p_1+2\,{\rm sign}(p_1)\Big]\;, 
\end{equation}
where ${\rm sign}(x)$ is the sign function with ${\rm sign}(0)=0\,$.

The above is a straightforward computation of the two-point functions (that does not require a regularization scheme), establishing that they are non-zero, as it should be 
for any two-point function of (anti-)hermitian operators. By the argument around (\ref{non-anomalousWard}) this proves that the Ward identities are violated, 
hence establishing  the presence  of an anomaly. 
However, it is not easy to interpret the corresponding effective 
action and hence to compute its (anomalous) transformation. We therefore turn to the more conventional Feynman diagram computation that does require a regularization scheme. 
As is customary also for chiral fermions we change gears by doing the computation on the plane, as opposed to the cylinder, which has the advantage that 
one regains Lorentz invariance. According to general lore, an anomaly does not depend on the topology \cite{Bilal:2008qx}, and so this should not affect the invariant result.  

We shall thus consider the action \eqref{ONFJ} on the plane and couple it to two-dimensional gauge fields $A_\alpha^{ab}\,$, promoting $SO(d)$ to a local symmetry:
\begin{equation}\label{actiongauged}
 S[\phi, A]=\frac{1}{2\pi\alpha'}\int d^2x \,D_1\phi^a\,D_-\phi_a\;,   
\end{equation}
where $D_\alpha \phi^a:=\del_\alpha\phi^a+A_\alpha^{ab}\,\phi_b$ and under a local $SO(d)$ rotation $\delta_\lambda A_\alpha^{ab}=-D_\alpha\lambda^{ab}\,$. In order to keep track of index contractions, it is useful to introduce the matrix
\begin{equation}
g^{\alpha\beta}:=\begin{pmatrix}0&1\\1&0
\end{pmatrix}\;,\quad{\rm where}\; \alpha=(-,1)\;,
\end{equation}
so that the above action can be recast in the form
\begin{equation}
S=\frac{1}{4\pi\alpha'}\int d^2x \,g^{\alpha\beta}\,D_\alpha\phi^a\,D_\beta\phi_a    \;,
\end{equation}
that allows to write the cubic and quartic vertices as
\begin{equation}
S_3=-\frac{1}{2\pi\alpha'}\int d^2x\,g^{\alpha\beta}A_{\alpha}^{ab}\,\phi_a\del_\beta\phi_b   \;,\quad S_4=\frac{1}{4\pi\alpha'}\int d^2x\,g^{\alpha\beta}A_\alpha^{ca}A_{\beta\,c}{}^b\,\phi_a\phi_b\;. 
\end{equation}
We define the one-loop effective action for $A_\alpha^{ab}$ by
\begin{equation}
e^{iW[A]}=Z^{-1}\int D\phi\,e^{iS[\phi, A]}=\left\langle e^{i(S_3+S_4)}\right\rangle  \;,  
\end{equation}
where $Z$ is the free $\phi-$path integral normalization and we denote averages by $\langle...\rangle\,$. We focus on the quadratic part of $W[A]\,$, that is given by
\begin{equation}
W_2[A]=\frac{i}{2}\,\langle S_3^2\rangle_{\rm conn.}+\langle S_4\rangle\;\sim\;\WEFF   \;.  
\end{equation}
By using the propagator
\begin{equation}
\langle\phi^a(x)\,\phi^b(y)\rangle=\frac{i\alpha'}{4\pi}\,\delta^{ab}\int d^2k\,\frac{e^{ik\cdot (x-y)}}{k_-k_1+i\epsilon}    
\end{equation}
and the Fourier representation
\begin{equation}
A_\alpha^{ab}(x)=\int d^2k\,e^{ik\cdot x}A_\alpha^{ab}(k)    \end{equation}
the effective action $W_2[A]$ can be written in the form
\begin{equation}
W_2[A]=-\frac{i}{4}\,\int d^2p\,g^{\alpha\gamma}g^{\beta\delta}\,A_\gamma^{ab}(p)\,\Pi_{\alpha\beta}(p)\,A_{\delta\,ab}(-p)\;.
\end{equation}
The polarization tensor $\Pi_{\alpha\beta}(p)$ reads
\begin{equation}\label{Polarization}
\Pi_{\alpha\beta}(p)=-\frac12\,\int dk_-dk_1\frac{(2k_\alpha-p_\alpha)(2k_\beta-p_\beta)-2\,g_{\alpha\beta}\big[k_-k_1+(p_--k_-)(p_1-k_1)\big]}{k_1(p_1-k_1)(k_-+\tfrac{i\epsilon}{k_1})(k_--p_--\tfrac{i\epsilon}{p_1-k_1})} \;, \end{equation}
where the integrand has manifest symmetry in the exchange $k_\alpha\leftrightarrow(p_\alpha-k_\alpha)\,$.  

We regularize the integral over $k_-$ by inserting a factor $e^{i\delta\,k_-k_1}\,$. This allows to compute the integral by closing the contour in the complex $k_--$plane. Although this regulator is not Lorentz invariant, it does not affect the anomaly. The remaining integral over $k_1$ suffers from infrared divergences, appearing as $\int_0^{|p_1|}\frac{dk_1}{k_1}\,$. This is to be expected from a two-dimensional massless scalar on the infinite plane, while this divergence disappears on the cylinder, due to the discrete spatial momentum. We will thus regulate the $k_1-$integral by introducing an infrared cutoff $\mu\,$, substituting $\int_0^{|p_1|}$ with $\int_\mu^{|p_1|}\,$. The various components of the polarization tensor read
\begin{equation}
\begin{split}
\Pi_{11}(p)&=-2\pi i\,\frac{p_1}{p_-}\,\big[\log(|p_1|/\mu)-2\big] \;,\\
\Pi_{--}(p)&=-2\pi i\,\frac{p_-}{p_1}\,\log(|p_1|/\mu)\;,\quad \Pi_{-1}(p)=2\pi i\,\log(|p_1|/\mu)\;,
\end{split}    
\end{equation}
yielding the effective action
\begin{equation}\label{W2}
\begin{split}
W_2[A]&=-\frac{\pi}{2}\int d^2p\,\Big\{A_-^{ab}(p)\,\frac{p_1}{p_-}\,\big[\log(|p_1|/\mu)-2\big]A_{-\,ab}(-p)\\
&\hspace{22mm}+A_1^{ab}(p)\,\frac{p_-}{p_1}\,\log(|p_1|/\mu)\,A_{1\,ab}(-p)-2\,A_1^{ab}(p)\,\log(|p_1|/\mu)\,A_{-\,ab}(-p)\Big\} \;.   
\end{split}    
\end{equation}
To recast the above expression in the more familiar lightcone basis one can substitute $A_1^{ab}=A_+^{ab}-A_-^{ab}$ and similarly $p_1=p_+-p_-\,$. Before computing the gauge variation of \eqref{W2}, we shall notice that the term $A_-^{ab}\frac{p_1}{p_-}A_{-\,ab}$ differs from $A_-^{ab}\frac{p_+}{p_-}A_{-\,ab}$ by a purely local term, implying that one can use the latter to compute the anomaly.

To lowest order in $A_\alpha^{ab}$ one has $\delta_\lambda A_\alpha^{ab}(p)=-ip_\alpha \lambda^{ab}(p)\,$, finally yielding the anomalous variation
\begin{equation}\label{anomaly}
\delta_\lambda W_2[A]=-2\pi i\int d^2p\,\lambda^{ab}(p)\,p_+\,A_{-\,ab}(-p) +{\cal O}(A^2)  \end{equation}
that is Lorentz invariant and structurally the same as the one due to a chiral fermion.
As usual, in order to establish that \eqref{anomaly} is a genuine anomaly, one has to consider adding to the effective action all possible local counterterms. Since \eqref{anomaly} is Lorentz invariant, the only local counterterm that can change it is
\begin{equation}
\Delta W_2[A]=\alpha\int d^2p\,A_+^{ab}(p)\,A_{-\,ab}(-p)\;,    
\end{equation}
leading to
\begin{equation}
\delta_\lambda(W_2+\Delta W_2)=-i\int d^2p\,\lambda^{ab}(p)\,\Big[(2\pi+\alpha)\,p_+A_{-\,ab}(-p)+\alpha\,p_-A_{+\,ab}(-p)\Big]\;.
\end{equation}
The above result shows that no value of $\alpha$ can make the effective action gauge invariant, thus establishing that \eqref{anomaly} is a genuine anomaly. We shall choose the value $\alpha=-\pi\,$, in order to have a purely parity-violating anomalous variation. Defining
\begin{equation}
W_2^{\rm eff}[A]:=W_2[A]-\pi\int d^2p\,A_+^{ab}(p)\,A_{-\,ab}(-p) \;,   
\end{equation}
with $W_2[A]$ given by \eqref{W2}, we finally obtain, using $\epsilon^{+-}=2\,$,
\begin{equation}\label{consistentanomaly}
\begin{split}
\delta_\lambda W_2^{\rm eff}[A]&= -i\pi\int d^2p\,\lambda^{ab}(p)\,\Big[p_+A_{-\,ab}(-p)-p_-A_{+\,ab}(-p)\Big]\\
&=\frac{1}{4\pi}\int d^2x\,\lambda_{ab}\,\Big[\del_+A_-^{ab}-\del_-A_+^{ab}\Big]=\frac{1}{8\pi}\int d^2x\,\lambda_{ab}\,\epsilon^{\alpha\beta}\del_\alpha A_\beta^{ab}\;.
\end{split}    
\end{equation}

\subsection{Green-Schwarz Deformation}

After computing the $SO(d)$ anomaly due to left-moving Floreanini-Jackiw bosons we can now apply the result to the worldsheet sigma-model \eqref{finalaction}. The action for the ``internal'' $Y-$sector \eqref{SYdecomposed} reads
\begin{equation}
\begin{split}
S_Y&=\frac{1}{2\pi\alpha'}\int d^2\sigma \,\Big[\big(\calD_1 Y^{\un A}+P_1{}^{\un A}{}_{\ov B}\,Y^{\ov B}\big)\big(\calD_- Y_{\un A}+P_{-\,\un A\ov C}\,Y^{\ov C}\big)\\
&\hspace{22mm}+\big(\calD_1 Y^{\ov A}+P_1{}^{\ov A}{}_{\un B}\,Y^{\un B}\big)\big(\calD_+ Y_{\ov A}+P_{+\,\ov A\un C}\,Y^{\un C}\big)\Big]\;,    
\end{split}    
\end{equation}
with $\calD_\alpha Y^{\un A}=\del_\alpha Y^{\un A}+Q_\alpha^{\un A\un B}\,Y_{\un B}\,$, $\calD_\alpha Y^{\ov A}=\del_\alpha Y^{\ov A}+Q_\alpha^{\ov A\ov B}\,Y_{\ov B}$ and we recall that $Q_\alpha=\del_\alpha X^\mu\,Q_\mu$ and $P_\alpha=\del_\alpha X^\mu\,P_\mu$ are the pullbacks of the background fields.

Our goal is to investigate the one-loop effective action (still depending on the $X^\mu$ worldsheet fields and background fields $Q_\mu$ and $P_\mu$) generated by integrating out the internal fields $(Y^{\un A}, Y^{\ov A})\,$. We first focus on the case $P_\alpha^{\un A\ov B}=0\,$, where one can see that the above action reduces to the sum
of a left-moving FJ action \eqref{actiongauged} with gauged $SO(d)_L$ and an analogous right-moving action with gauged $SO(d)_R\,$. By just replacing $\phi^a\to Y^{\un A}\,$, $A_\alpha^{ab}\to Q_\alpha^{\un A\un B}$ one obtains an effective action $W_2[Q, \bar Q]$ whose anomalous $SO(d)_L\times SO(d)_R$ variation is given, to lowest order, by
\begin{equation}\label{anomalyQ}
\delta_{\lambda,\bar\lambda}W_2[Q,\bar Q]=\frac{1}{8\pi}\int d^2x\,\lambda_{\un A\un B}\,\epsilon^{\alpha\beta}\del_\alpha Q_\beta^{\un A\un B}-\frac{1}{8\pi}\int d^2x\,\lambda_{\ov A\ov B}\,\epsilon^{\alpha\beta}\del_\alpha Q_\beta^{\ov A\ov B}\;,     
\end{equation}
where the right-moving contribution can be obtained by a computation analogous to the one presented in the previous subsection.

We can now examine the effect of the $P_\alpha$ tensor on the full effective action $W_2[Q,\bar Q, P]\,$. Due to the orthogonality of the two $SO(d)$ groups, the only contribution of $P_\alpha$ to the quadratic effective action $W_2$ has to be of the form 
\begin{equation}\label{WP}
\int d^2k\,P_\alpha^{\un A\ov B}(k)\,G^{\alpha\beta}(k)\,P_{\beta\,\un A\ov B}(-k)\;.    
\end{equation}
Since the gauge transformation of the $P-$tensor is $\delta_\lambda P_{\mu\,\un A\ov B}=\lambda_{\un A}{}^{\un C}\,P_{\mu\,\un C\ov B}+\lambda_{\ov B}{}^{\ov C}\,P_{\mu\,\un A\ov C}\,$, the variation of \eqref{WP}, if non-vanishing, cannot contribute to linear order in the background fields. This shows that, to lowest order in the fields, the anomalous variation of $W[Q,\bar Q, P]$ is given by \eqref{anomalyQ}, that can be written in form language as
\begin{equation}\label{finalanomaly}
 \delta_{\lambda,\bar\lambda}W[Q,\bar Q, P]=\frac{1}{8\pi}\int\,{\rm tr}\Big(d\lambda\wedge Q\Big)-\frac{1}{8\pi}\int\,{\rm tr}\Big(d\bar\lambda\wedge \bar Q\Big) \;.  
\end{equation}
Since the above result already satisfies the Wess-Zumino consistency conditions $[\delta_{\lambda_1},\delta_{\lambda_2}]W=\delta_{[\lambda_2,\lambda_1]}W\,$, \eqref{finalanomaly} does not receive higher order contributions in $Q$ and $P\,$, and gives the full anomaly.
Let us mention that in cosmological settings, where the ``external'' coordinates $X^\mu$ reduce to time $X^0=t\,$, one has $dQ^{ab}=d\sigma^\alpha\!\!\wedge d\sigma^\beta\,\del_\alpha t\del_\beta t\,\del_t Q^{ab}_t\equiv0$ and the anomaly is not present.

The above anomalous variation, if not canceled, implies that gauge-equivalent background fields (from the target space perspective) lead to inequivalent worldsheet sigma-models, which is not acceptable. Fortunately, in the same spirit of the original Green-Schwarz mechanism \cite{Hull:1985jv}, the anomaly \eqref{finalanomaly} can be canceled by postulating a suitable transformation for the $B-$field. We recall from \eqref{finalaction}
that the action involving the Kalb-Ramond field reads
\begin{equation}
S_B=-\frac{1}{4\pi\alpha'}\int d^2x\,\epsilon^{\alpha\beta}\,B_{\alpha\beta}\;,
\end{equation}
where $B_{\alpha\beta}$ denotes the pullback of $B_{\mu\nu}\,$. At this point, simple inspection of \eqref{finalanomaly} determines that the anomaly can be canceled by assigning to the $B-$field the transformation law
\begin{equation}\label{GS}
\delta_{\lambda,\bar\lambda}B= \frac{\alpha'}{2}\,{\rm tr}\Big(d\lambda\wedge Q\Big)-\frac{\alpha'}{2}\,{\rm tr}\Big(d\bar\lambda\wedge \bar Q\Big) \;.  
\end{equation}
Remarkably, this is exactly the transformation found in \cite{Eloy:2019hnl} from the low-energy target space analysis, thus showing that its emergence in the worldsheet theory stems from the anomalies of two-dimensional chiral bosons.

As a final comment, the form \eqref{finalanomaly} of the $SO(d)_L\times SO(d)_R$ anomaly shows that the diagonal $SO(d)$ subgroup remains unbroken. This also agrees with the analysis of \cite{Eloy:2019hnl} and is to be expected, since the diagonal $SO(d)$ is the geometric subgroup of $SO(d)_L\times SO(d)_R\,$.

\section{Conclusions and Outlook}

In this paper we have revisited the issue of making the T-duality group $O(d,d)$ a manifest symmetry of the worldsheet action of (bosonic) string theory. 
We have identified a consistent truncation with global $O(d,d,\mathbb{R})$ invariance,  in which the target space fields are independent of $d$ coordinates while the worldsheet scalars have zero (internal) momentum and winding. This truncation may be thought of as the zero-mass sector for a Kaluza-Klein compactification on a $d$-dimensional torus, 
but the topology is no longer relevant --- precisely because of the truncation to zero momentum and winding. 
As such, this worldsheet theory is applicable to any setting with $d$ abelian isometries, be they compact or not, in particular to cosmological backgrounds, as employed in 
\cite{Hohm:2019jgu}. 
We have displayed the proper manifestly $O(d,d,\mathbb{R})$ invariant worldsheet action that includes all target space 
fields that survive the truncation.
 
As the second main point of this paper we have shown that the $SO(d)_L\times SO(d)_R$ local frame transformations are anomalous, as to be expected 
given the presence of  chiral bosons. This suggests that a Green-Schwarz mechanism is needed in which the (external) B-field, which is a singlet 
in the classical theory, transforms non-trivially under these symmetries, in line with recent findings in the target space theory when higher oder $\alpha'$ 
corrections are included \cite{Eloy:2019hnl,Eloy:2020dko}.  This result has a direct bearing on any attempts to determine  the target space equations 
directly in $O(d,d,\mathbb{R})$ invariant form by computing the beta functions of a suitable $O(d,d,\mathbb{R})$ invariant worldsheet theory, a program that 
was initiated in \cite{Berman:2007vi,Berman:2007xn,Berman:2007yf}. It will then be important to revisit this program in light of the present results, in particular to develop precise computational rules that allow one, in principle, to determine the equations to arbitrary orders in $\alpha'$. In this respect, one of the main difficulties in using this formalism is the lack of manifest two-dimensional Lorentz invariance. It would then be interesting to investigate the proposal, made in \cite{Tseytlin:2}, of modifying the functional measure of the chiral bosons as $D\phi\to D\phi\,({\rm det}\,\del_1)^{1/2}\,$. This formally relates the path integral of the Floreanini-Jackiw bosons to the one of chiral fermions, which is manifestly Lorentz invariant.

A manifestly $O(d,d,\mathbb{R})$ invariant procedure to compute the beta functions  
may be particularly fruitful in the cosmological setting, which is significantly simplified since the external dimensions are reduced to (cosmic) time, 
and where a complete classification of all duality invariant $\alpha'$ corrections has been found recently \cite{Hohm:2019jgu}. 
It remains to fix a  finite number of free parameters at each order in $\alpha'$, and one may hope that this could eventually  be achieved by a worldsheet computation 
using the results given here.

Let us finally mention that while the general phenomenon for which we provide here a worldsheet interpretation was first discovered in 
double field theory \cite{Hohm:2013jaa,Hohm:2014xsa}, none of our findings depend directly on double field theory. They are a feature of a standard 
string theory formulation. 
Nevertheless, 
the most enticing extension of this framework would of course be to a full-fledged double field theory. 
In the truncation invoked here there is a clear separation of dimensions along which the fields may vary (external) and of dimensions along which the fields  are constant (internal), 
with the $O(d,d,\mathbb{R})$ acting exclusively on the latter.  
This truncation is explicitly $O(d,d,\mathbb{R})$ invariant to all orders in $\alpha'$, as follows by general arguments \cite{Sen:1991zi} and explicit computations \cite{Meissner:1996sa,Hohm:2019jgu}, and so there should be a worldsheet CFT construction giving these target space equations. 
However, a genuine double field theory would go beyond this by having fields that in addition depend on doubled internal coordinates, corresponding to the 
scalar fields $Y^M$, subject to the level-matching constraint (that now does assume a torus background) and obeying a novel algebra \cite{Hull:2009mi}. 
The results obtained here may help to illuminate some issues that arise when trying to define this theory explicitly.

\section*{Acknowledgements}

We would like to thank Fiorenzo Bastianelli, Chris Blair, Ashoke Sen and Arkady Tseytlin  for helpful discussions. 
This work is supported by the ERC Consolidator Grant ``Symmetries and Cosmology".

\appendix

\section*{Appendix}

\section{Worldsheet Diffeomorphisms}\label{sec:diffeos}

Here we derive various non-standard  realizations of worldsheet diffeomorphisms, which in our formulation are not manifest. 
In particular, we show that in the Hamiltonian picture they are generated by the Virasoro constraints, as to be expected. 
More importantly, we also show that the $\widetilde{Y}_i$, introduced by a non-local field redefinition $2\pi\alpha'P_i=\del_\sigma\widetilde Y_i$, 
admit local diffeomorphism transformations. Finally, we show that diffeomorphism invariance, despite its non-standard realization, implies 
energy-momentum conservation in the standard  form.

\subsection*{Equivalence between Lagrangian and Hamiltonian diffeomorphisms}

Let us start from the sigma model action \eqref{Polyakov NOT constant}, \emph{i.e.}
\begin{equation}\label{Polyakov NOT constant1}
S_L=-\frac{1}{4\pi\alpha'}\int d^2\sigma\big[\sqrt{-h}h^{\alpha\beta}\,\hat G_{\hat\mu\hat\nu}(X)+\epsilon^{\alpha\beta}\,\hat B_{\hat\mu\hat\nu}(X)\big]\del_\alpha\hat X^{\hat\mu}\del_\beta\hat X^{\hat\nu}  \;, 
\end{equation}
where here and in the following the subscript $L$ refers to `Lagrangian', as opposed to $H$ which will refer to `Hamiltonian'.  
Diffeomorphism and Weyl transformations  take the familiar form 
\begin{equation}
\begin{split}
\delta_L h_{\alpha\beta}&=\xi^\lambda\del_\lambda h_{\alpha\beta}+2\,\del_{(\alpha}\xi^\lambda\,h_{\beta)\lambda}+2\omega h_{\alpha\beta}\;,\\
\delta_L\hat X^{\hat\mu}&=\xi^\alpha\del_\alpha \hat X^{\hat\mu}\;.
\end{split}
\end{equation}
Using the parametrization \eqref{eu} for the metric one finds 
\begin{equation}\label{deltaLeu}
\begin{split}
\delta_L\Omega&=\del_\alpha(\Omega\,\xi^\alpha)+2\omega\,\Omega\;,\\
\delta_Le&=\xi^\alpha\del_\alpha e+e\,\big[\del_\tau\xi^\tau-\del_\sigma\xi^\sigma-2\,u\,\del_\sigma\xi^\tau\big]\;,\\
\delta_Lu&=\xi^\alpha\del_\alpha u+\del_\tau\xi^\sigma+u\,\big[\del_\tau\xi^\tau-\del_\sigma\xi^\sigma\big]-\del_\sigma\xi^\tau\,\big[u^2+e^2\big]\;,
\end{split}    
\end{equation}
which  is awkward, since the basis $(e,u,\Omega)$ is adapted to the Hamiltonian formulation.

We will now determine the Hamiltonian form of the diffeomorphisms, which are generated  via Poisson brackets from the Virasoro constraints. 
The Hamiltonian action associated to \eqref{Polyakov NOT constant1} was given in (\ref{Hamiltonaction}), 
except that now we would have to replace all indices by hatted indices, referring to the totality  of internal and external components. 
However, here we will not be concerned with the split into ``external'' $X^\mu$ and ``internal'' $Y^i\,$, and so  in order not to overburden the notation, we shall drop all hats from our formulas in what follows. In particular, $X^\mu$ and $P_\mu$ stand for  $(n+d)$-dimensional phase space variables, $G_{\mu\nu}$ and $B_{\mu\nu}$ denote the spacetime metric and $B$-field in $(n+d)$ dimensions and, finally, capital indices $M=1,\ldots ,2(d+n)$ denote $O(d+n,d+n)$ tensors. 

The fundamental Poisson brackets are 
\begin{equation}
\big\{X^\mu(\sigma_1),\,P_\nu(\sigma_2)\big\}=\delta^\mu_\nu\,\delta(\sigma_1-\sigma_2) \,.    
\end{equation}
The simple form of the constraints in terms of $O(D,D)$ quantities in (\ref{OddConstr}) suggests to use the covariant Poisson brackets
\begin{equation}\label{ZZbracket}
\big\{Z^M(\sigma_1), Z^N(\sigma_2)\big\}=2\pi\alpha'\,\eta^{MN}\,\del_{\sigma_1}\delta(\sigma_1-\sigma_2)\;, 
\end{equation}
along with
\begin{equation}\label{ZfofXbracket}
\big\{Z_M(\sigma_1),\Phi(X(\sigma_2))\big\}=-2\pi\alpha'\,\del_M\Phi(X)\,\delta(\sigma_1-\sigma_2)    
\end{equation}
for $X$-dependent fields, where it is understood that $\tilde\del^\mu\Phi(X)\equiv0\,$. Using \eqref{ZZbracket} and \eqref{ZfofXbracket} it is indeed simple to compute the local constraint algebra:
\begin{equation}\label{localdiff2}
\begin{split}
\big\{\cN(\sigma_1), \cN(\sigma_2)\big\}&=\tfrac{1}{2\pi\alpha'}\,Z^M(\sigma_1)Z_M(\sigma_2)\,\del_{\sigma_1}\delta(\sigma_1-\sigma_2)\;,\\[2mm] 
\big\{\cN(\sigma_1), \cH(\sigma_2)\big\}&=\tfrac{1}{2\pi\alpha'}\,Z^M(\sigma_1)Z^N(\sigma_2)\,\cH_{MN}(\sigma_2)\,\del_{\sigma_1}\delta(\sigma_1-\sigma_2)\\
&\hspace{5mm}-\tfrac{1}{4\pi\alpha'}\,Z^P(\sigma_1)\del_P\cH_{MN}(\sigma_1)\,Z^M(\sigma_1)Z^N(\sigma_1)\,\delta(\sigma_1-\sigma_2)\;,\\[2mm]
\big\{\cH(\sigma_1),\cH(\sigma_2)\big\}&=\tfrac{1}{2\pi\alpha'}\,Z^M(\sigma_1)\cH_{MP}(\sigma_1)\,\cH^P{}_N(\sigma_2)Z^N(\sigma_2)\,\del_{\sigma_1}\delta(\sigma_1-\sigma_2)\;,
\end{split}    
\end{equation}
where we can reduce $Z^P\del_P\cH_{MN}=\del_\sigma X^\mu\del_\mu\cH_{MN}=\del_\sigma\cH_{MN}$. 
Given the local form \eqref{localdiff2}, the constraint algebra, adopting the formalism in Refs. \cite{Blair:2013noa} and \cite{Townsend:2019koy}, is most easily read in terms of the smeared constraints
\begin{equation}\label{smearedconstraints}
N(\alpha):=\int_0^{2\pi} d\sigma\,\alpha(\sigma)\,\cN(\sigma)\;,\quad H(\epsilon):=\int_0^{2\pi} d\sigma\,\epsilon(\sigma)\,\cH(\sigma)\;,    
\end{equation}
for which the first-class property becomes apparent:
\begin{equation}\label{smearedalgebra}
\begin{split}
\big\{N(\alpha_1),N(\alpha_2)\big\}&=N([\alpha_1,\alpha_2]) \;,\\   
\big\{N(\alpha),H(\epsilon)\big\}&=H([\alpha,\epsilon]) \;,\\
\big\{H(\epsilon_1),H(\epsilon_2)\big\}&=N([\epsilon_1,\epsilon_2]) \;,
\end{split}    
\end{equation}
with square brackets denoting one-dimensional Lie brackets, \emph{i.e.}
\begin{equation}
[\alpha_1,\alpha_2]:=\alpha_1 \del_\sigma\alpha_2-\alpha_2 \del_\sigma\alpha_1\;.    
\end{equation}
One can choose a diagonal basis  for $\cN$ and $\cH\,$ by using the projectors $\Pi^\pm_{MN}:=\tfrac12\,\big(\eta_{MN}\pm\cH_{MN}\big)$ 
(which can be quickly verified to be projectors thanks to $\cH_{MN}$ being an $O(D,D)$ element): 
\begin{equation}\label{diagonalHs}
\cH_\pm:=\frac12\,\big(\cN\pm\cH\big)=
\frac{1}{4\pi\alpha'}\,\Pi^\pm_{MN}\,Z^MZ^N \;.  \end{equation}
The Hamiltonian algebra \eqref{smearedalgebra} then takes the manifest $\frak{diff}_1\oplus\frak{diff}_1\,$ form: 
\begin{equation}
\big\{H_\pm(\alpha_1),H_\pm(\alpha_2)\big\}=H_\pm([\alpha_1,\alpha_2])\;,\quad\big\{H_+(\alpha),H_-(\beta)\big\}=0\;.   \end{equation}

The smeared constraints \eqref{smearedconstraints} can be used to determine the Hamiltonian gauge transformations of the phase space fields. Taking Poisson brackets one finds
\begin{equation}\label{deltaHXP}
\begin{split}
\delta_HX^\mu&=\big\{X^\mu, N(\alpha)+H(\epsilon)\big\}=\alpha\,\del_\sigma X^\mu+\epsilon\,\cH^\mu{}_N\,Z^N\;,\\
\delta_HP_\mu&=\big\{P_\mu, N(\alpha)+H(\epsilon)\big\}=\del_\sigma\big[\alpha\,P_\mu+\tfrac{1}{2\pi\alpha'}\,\epsilon\,\cH_{\mu N}\,Z^N\big]-\tfrac{1}{4\pi\alpha'}\,\epsilon\,\del_\mu\cH_{MN}\,Z^MZ^N\;,
\end{split}    
\end{equation}
that are the Hamiltonian version of $\frak{diff}_2$ transformations. Finally, requiring invariance of the Hamiltonian action  under the above local transformations determines the transformation law of the corresponding Hamiltonian gauge fields $e$ and $u\,$:
\begin{equation}\label{deltaHeu}
\delta_He=\del_\tau\epsilon-[u,\epsilon]-[e,\alpha]\;,\quad\delta_Hu=\del_\tau\alpha-[u,\alpha]-[e,\epsilon]\;,    
\end{equation}
where again square brackets denote one-dimensional Lie brackets. 

The transformation laws \eqref{deltaHXP} and \eqref{deltaHeu} have to be compared with their Lagrangian counterpart \eqref{deltaLeu} and $\delta_LX^\mu=\xi^\alpha\del_\alpha X^\mu\,$. Defining the Lagrangian transformation $\delta_LP_\mu$ for momenta may seem counter intuitive, but it is simply determined by considering the transformation of $P^{\rm lag}_\mu\equiv\frac{\del\cL}{\del\dot X^\mu}$ as a given function of Lagrangian variables, \emph{i.e.}
\begin{equation}
P_\mu^{\rm lag}=\frac{1}{2\pi\alpha'}\Big[\tfrac{1}{e}\,G_{\mu\nu}(\del_\tau X^\nu-u\,\del_\sigma X^\nu)+B_{\mu\nu}\,\del_\sigma X^\nu\Big] \;,   
\end{equation}
that, obviously, is just the on-shell value of the Hamiltonian momentum. Using the transformation law of $P_\mu^{\rm lag}$ to define $\delta_LP_\mu$ 
ensures that such transformations commute with integrating out momenta.

The simplest way to determine the transformation law of $P_\mu^{\rm lag}$ is to view it as the $\tau$-component of the two-dimensional vector density
\begin{equation}
\pi_\mu^\alpha:=-\frac{1}{2\pi\alpha'}\,\Big[\sqrt{-h}h^{\alpha\beta}\,G_{\mu\nu}\,\del_\beta X^\nu+\epsilon^{\alpha\beta}\,B_{\mu\nu}\,\del_\beta X^\nu\Big] \;,   
\end{equation}
that transforms as
\begin{equation}
\delta_L\pi_\mu^\alpha=\del_\beta (\xi^\beta\,\pi^\alpha_\mu)-\del_\beta\xi^\alpha\,\pi^\beta_\mu \;,   
\end{equation}
thus yielding
\begin{equation}\label{deltaLXP}
\begin{split}
\delta_LX^\mu&=\xi^\alpha\del_\alpha X^\mu\;,\\
\delta_LP_\mu&=\xi^\tau\,\del_\tau P_\mu+\del_\sigma(\xi^\sigma\,P_\mu)+\del_\sigma\xi^\tau\,\big(u\,P_\mu+\tfrac{1}{2\pi\alpha'}\,e\,\cH_{\mu N}\,Z^N\big)\;. \end{split}   
\end{equation}
At this point, the final ingredient to prove equivalence of the Lagrangian $\frak{diff}_2$ transformations \eqref{deltaLXP} with the canonical ones \eqref{deltaHXP}, are the Hamiltonian field equations 
\begin{equation}
\begin{split}
\frac{\delta S_H}{\delta P_\mu}&=\del_\tau X^\mu-u\,\del_\sigma X^\mu-e\,\cH^\mu{}_N\,Z^N\;,\\
\frac{\delta S_H}{\delta X^\mu}&=-\del_\tau P_\mu+\del_\sigma\big[u\,P_\mu+\tfrac{1}{2\pi\alpha'}\,e\,\cH_{\mu N}\,Z^N\big]-\frac{e}{4\pi\alpha'}\,\del_\mu\cH_{MN}\,Z^MZ^N\;.
\end{split}    
\end{equation}
Inspection of the transformation laws \eqref{deltaHXP} and \eqref{deltaLXP} allows us to determine the relation between the canonical gauge parameters $(\alpha,\epsilon)$ and the geometric vector field $\xi^\alpha\,$, namely
\begin{equation}\label{epsilonalphaxi}
\epsilon=e\,\xi^\tau\;,\quad \alpha=\xi^\sigma+u\,\xi^\tau  \;.  
\end{equation}
With the above redefinition we can finally establish the explicit equivalence between two-dimensional diffeomorphisms and canonical gauge transformations as
\begin{equation}
\begin{split}
&\delta_HX^\mu=\delta_LX^\mu-\xi^\tau\,\frac{\delta S_H}{\delta P_\mu}\;,\quad \delta_HP_\mu=\delta_LP_\mu+\xi^\tau\,\frac{\delta S_H}{\delta X^\mu}\;,\\
&\delta_He=\delta_Le\;,\quad \delta_Hu=\delta_Lu\;.
\end{split}    
\end{equation}
The extra terms in $\delta_HX^\mu$ and $\delta_HP_\mu$ are indeed of the trivial form $\delta\varphi^i=\mu^{ij}\frac{\delta S}{\delta\varphi^j}$ with $\mu^{ij}$ antisymmetric. This kind of local transformations is not related to any genuine gauge redundancy, and can be safely ignored.

\subsection*{Locality and $O(d,d)$ invariance of diffeomorphisms}

After proving equivalence of Lagrangian diffeomorphisms \eqref{deltaLXP} and Hamiltonian gauge transformations \eqref{deltaHXP} for the general sigma model in $(n+d)$ dimensions, we shall now study the split $\hat X^{\hat\mu}=(X^\mu,Y^i)$ between external and internal sectors.

First of all, let us reinstate the original notation, with hatted symbols denoting $(n+d)$-dimensional fields, and rewrite the Lagrangian and Hamiltonian transformations: 
\begin{equation}\label{deltaLXPhats}
\begin{split}
\delta_L\hat X^{\hat\mu}&=\xi^\alpha\del_\alpha \hat{X}^{\hat\mu}\;,\\
\delta_L\hat P_{\hat\mu}&=\xi^\tau\,\del_\tau \hat P_{\hat\mu}+\del_\sigma(\xi^\sigma\,\hat P_{\hat\mu})+\del_\sigma\xi^\tau\,\big(u\,\hat P_{\hat\mu}+\tfrac{1}{2\pi\alpha'}\,e\,\hat\cH_{\hat\mu \hat N}\,\hat Z^{\hat N}\big)\;, \end{split}   
\end{equation}
and
\begin{equation}\label{deltaHXPhats}
\begin{split}
\delta_H\hat X^{\hat\mu}&=\alpha\,\del_\sigma \hat X^{\hat\mu}+\epsilon\,\hat\cH^{\hat\mu}{}_{\hat N}\,\hat Z^{\hat N}\;,\\
\delta_H\hat P_{\hat\mu}&=\del_\sigma\big[\alpha\,\hat P_{\hat\mu}+\tfrac{1}{2\pi\alpha'}\,\epsilon\,\hat\cH_{\hat\mu \hat N}\,\hat Z^{\hat N}\big]-\tfrac{1}{4\pi\alpha'}\,\epsilon\,\del_{\hat\mu}\hat\cH_{\hat M\hat N}\,\hat Z^{\hat M}\hat Z^{\hat N}\;.
\end{split}    
\end{equation}
Let us also remind that the $O(d+n,d+n)$ vector $\hat Z^{\hat M}$ is given by
\begin{equation}\label{Zhat}
\hat Z^{\hat M}=\begin{pmatrix}
\del_\sigma\hat X^{\hat\mu}\\2\pi\alpha'\,\hat P_{\hat\mu}
\end{pmatrix}    \;,
\end{equation}
and the $O(d+n,d+n)$ generalized metric is defined in terms of $\hat G_{\hat\mu\hat\nu}$ and $\hat B_{\hat\mu\hat\nu}\,$. 

Upon splitting the phase space variables as $\hat X^{\hat\mu}=(X^\mu,Y^i)$ and $\hat P_{\hat\mu}=(P_\mu,P_i)\,$, we recall that our action \eqref{finalaction} is purely Lagrangian in the non-compact sector. The diffeomorphism transformations thus act as usual: $\delta X^\mu=\xi^\alpha\del_\alpha X^\mu\,$, the on-shell momentum $P_\mu$ is given by
\begin{equation}\label{Pmuonshell2}
2\pi\alpha'\,P_\mu=e^{-1}g_{\mu\nu}\,\mathring{X}^\nu+\cB_{\mu\nu}\,\del_\sigma X^\nu+\cA_\mu{}^M\,Z_M  \;,\qquad \mathring{X}^\mu:=\del_\tau X^\mu-u\,\del_\sigma X^\mu  \;,
\end{equation}
according to \eqref{Pmuonshell}, and no further investigation is required.

For the internal $(Y^i,P_i)$ sector, we choose the Hamiltonian form \eqref{deltaHXPhats} over the Lagrangian one \eqref{deltaLXPhats}, a choice that we will motivate at the end of this section. According to \eqref{deltaHXPhats}, the transformations for the phase space variables $Y^i$ and $P_i$ are given by
\begin{equation}\label{deltaHYP}
\begin{split}
\delta Y^i&=\alpha\,\del_\sigma Y^i+\epsilon\,\hat\cH^{i}{}_{\hat N}\,\hat Z^{\hat N}\;,\\
\delta P_i&=\del_\sigma\big[\alpha\,P_i+\tfrac{1}{2\pi\alpha'}\,\epsilon\,\hat\cH_{i \hat N}\,\hat Z^{\hat N}\big]\;,
\end{split}    
\end{equation}
where, crucially, the last term  of \eqref{deltaHXPhats} vanishes in the transformation of $P_i\,$, thanks to $\del_i\Phi(X)=0$ for any spacetime field. This last fact implies the most important property we were after: $P_i$ transforms as a total $\sigma$-derivative under diffeomorphisms.\footnote{This also implies that the center of mass truncation $p_{i\,0}=0$ is diffeomorphism invariant.} This allows us to implement the field redefinition $2\pi\alpha'P_i=\del_\sigma\widetilde Y_i$ without introducing non-localities in the transformations. Indeed,  consistently with \eqref{deltaHYP}, we can write
\begin{equation}\label{deltaHYtildeY}
\begin{split}
\delta Y^i&=\alpha\,\del_\sigma Y^i+\epsilon\,\hat\cH^{i}{}_{\hat N}\,\hat Z^{\hat N}\;,\\
\delta \widetilde Y_i&=\alpha\,\del_\sigma\widetilde Y_i+\epsilon\,\hat\cH_{i \hat N}\,\hat Z^{\hat N}\;.
\end{split}    
\end{equation}
At this point, $O(d,d)$ invariance may look manifest. However, despite the simple-looking form, the decomposition of the $O(d+n,d+n)$ generalized metric in terms of $O(d,d)$ covariant $n$-dimensional fields is somewhat involved, see \cite{Hohm:2014sxa}. 
Upon using the Kaluza-Klein decomposition \eqref{KKdictionaryG}, \eqref{KKdictionaryB}, as well as \eqref{Zhat} and \eqref{Pmuonshell2} we find, after a straightforward but tedious computation
\begin{equation}
\begin{split}
\hat\cH^{i}{}_{\hat N}\,\hat Z^{\hat N}&=\cH^i{}_M\,D_\sigma Y^M-e^{-1}A_\mu^i(\del_\tau X^\mu-u\,\del_\sigma X^\mu)\;,\\
\hat\cH_{i\hat N}\,\hat Z^{\hat N}&=\cH_{iM}\,D_\sigma Y^M-e^{-1}\tilde A_{\mu\,i}(\del_\tau X^\mu-u\,\del_\sigma X^\mu)\;.
\end{split}    
\end{equation}
This establishes the diffeomorphism transformations of the double coordinates $Y^M$
in a manifestly local and $O(d,d)$ covariant form:
\begin{equation}
\delta Y^M=\alpha\,\del_\sigma Y^M+\epsilon\,\Big[\cH^{MN}\,D_\sigma Y_N-e^{-1}\cA_\mu{}^M(\del_\tau X^\mu-u\,\del_\sigma X^\mu)\Big] \;.   
\end{equation}
Recalling the relation \eqref{epsilonalphaxi} between the Hamiltonian gauge parameters $(\alpha,\epsilon)$ and the $\frak{diff}_2$ vector $\xi^\alpha\,$, it is possible to rewrite the above transformation law in a  more illuminating form:
\begin{equation}\label{gooddeltaY}
\delta Y^M=\xi^\alpha\del_\alpha Y^M-\xi^\tau\,\Big[D_\tau Y^M-u\,D_\sigma Y^M-e\,\cH^{MN}\,D_\sigma Y_N\Big]    \;.
\end{equation}
The above transformation law reduces to the standard one, $\delta Y^M=\xi^\alpha\del_\alpha Y^M\,$, upon using the self-duality relation \eqref{dualityarbitrarygauge}. However, \eqref{gooddeltaY} provides the correct \textit{off-shell} $\frak{diff}_2$ transformation in the general case. 
To summarize, the action \eqref{finalaction} is invariant under the worldsheet diffeomorphisms
\begin{equation}\label{diff2trans}
\begin{split}
&\delta_\xi X^\mu=\xi^\alpha\del_\alpha X^\mu\;,\quad\delta_\xi h_{\alpha\beta}=\nabla_\alpha\xi_\beta+\nabla_\beta\xi_\alpha\;,\\    
&\delta_\xi Y^M=\xi^\alpha\del_\alpha Y^M-\xi^\tau\,\cD^M\;,
\end{split}    
\end{equation}
where we defined the ``self-duality vector''
\begin{equation}
\cD^M:=D_\tau Y^M-u\,D_\sigma Y^M-e\,\cH^{MN}\,D_\sigma Y_N\;,    
\end{equation}
and we recall that the transformation law \eqref{deltaLeu} of $e$ and $u$ is just determined by their definition
\begin{equation}
e=\frac{\sqrt{-h}}{h_{\sigma\sigma}}\;,\quad u=\frac{h_{\tau\sigma}}{h_{\sigma\sigma}} \;.   
\end{equation}
Invariance of \eqref{finalaction} under \eqref{diff2trans} is assured by the general reasoning leading to \eqref{gooddeltaY}, but it can also be checked directly by using
\begin{equation}\label{deltaduality}
\begin{split}
\delta_\xi(D_\alpha Y^M)&=\cL_\xi(D_\alpha Y^M)-\del_\alpha(\xi^\tau\,\cD^M)\;,\\
\delta_\xi \cD^M&=\xi^\alpha\del_\alpha\cD^M-\xi^\tau\,\Big[\del_\tau\cD^M-u\,\del_\sigma\cD^M-e\,\cH^{MN}\,\del_\sigma \cD_N\Big]\;.
\end{split}    
\end{equation}
The variations \eqref{deltaduality} can also be used to check that the algebra of diffeomorphisms closes off-shell, even with the extra term, according to the usual Lie bracket:
\begin{equation}\label{diff2algebra}
[\delta_{\xi_2},\delta_{\xi_1}]=\delta_{\xi_{12}}\;,\quad \xi_{12}^\alpha=\xi_1^\beta\del_\beta\xi_2^\alpha-\xi_2^\beta\del_\beta\xi_1^\alpha   \;. 
\end{equation}

The zero mode shift symmetry \eqref{Xisymmetry}
\begin{equation}
\delta_\Xi Y^M(\sigma,\tau)=\Xi^M(\tau)\;,\quad \delta_\Xi X^\mu(\sigma,\tau)=0\;,\quad \delta_\Xi h_{\alpha\beta}(\sigma,\tau)=0 \;,
\end{equation}
commutes with diffeomorphisms: $[\delta_\xi,\delta_\Xi]=0\,$. However, the gauge fixing condition \eqref{Y0equation} is not $\frak{diff}_2$ invariant off-shell. The easiest way to see this is to notice that \eqref{Y0equation} can be written as
\begin{equation}
\frac{1}{2\pi}\int_0^{2\pi}d\sigma\,\cD^M=0\;,    
\end{equation}
and its variation under a diffeomorphism is given by
\begin{equation}
\delta_\xi\left(\int_0^{2\pi}d\sigma\,\cD^M\right)=\int_0^{2\pi}d\sigma\,\Big[\xi^\sigma\del_\sigma\cD^M+\xi^\tau(u\,\del_\sigma\cD^M+e\,\cH^{MN}\,\del_\sigma \cD_N)\Big] \;,   
\end{equation}
that vanishes only on-shell, by noting that the $Y$ field equation \eqref{dsigmaduality} is just $\del_\sigma\cD^M=0\,$. This is not in contradiction with our claim, namely that the string theory described by
\eqref{finalaction} is \emph{classically} equivalent to the truncated sector of the original sigma model, meaning that the equivalence holds  at the level of the space of classical solutions.

\subsection*{Energy-momentum tensor and conformal symmetry}

We will now derive consequences of diffeomorphism invariance such as energy-momentum conservation. 
The diffeomorphism invariance of the action \eqref{finalaction} can be expressed as
\begin{equation}\label{diffinvariance}
\int d^2\sigma\,\Big[\frac{\delta S}{\delta\varphi^A}\delta_\xi\varphi^A+\frac{\delta S}{\delta e}\delta_\xi e+\frac{\delta S}{\delta u}\delta_\xi u\Big]=0\;,  
\end{equation}
where we grouped the ``matter fields'' as $\varphi^A:=(X^\mu,Y^M)\,$. 
By evaluating \eqref{diffinvariance} for on-shell configurations of the matter fields, \emph{i.e.}~$\frac{\delta S}{\delta\varphi^A}=0\,$, one obtains
\begin{equation}\label{Tconsalmost}
\int d^2\sigma\Big[\cH\,\big(\del_\tau\epsilon-u\,\del_\sigma\epsilon-e\,\del_\sigma\alpha+\del_\sigma e\,\alpha\big)+\cN\,\big(\del_\tau\alpha-u\,\del_\sigma\alpha-e\,\del_\sigma\epsilon+\del_\sigma e\,\epsilon\big)\Big]  =0\;, 
\end{equation}
where we used \eqref{deltaHeu}. The functions $\cH=-\frac{\delta S}{\delta e}$ and $\cN=-\frac{\delta S}{\delta u}$ are given by 
\begin{equation}
\begin{split}
&\cH=\frac{1}{4\pi\alpha'}\,\Big[e^{-2}\,g_{\mu\nu}\,\mathring{X}^\mu\mathring{X}^\nu+g_{\mu\nu}\,\del_\sigma X^\mu\del_\sigma X^\nu+\cH_{MN}\,D_\sigma Y^M D_\sigma Y^N\Big]\;,\\
&\cN=\frac{1}{4\pi\alpha'}\,\Big[2\,e^{-1}\,g_{\mu\nu}\,\mathring{X}^\mu\del_\sigma X^\nu+D_\sigma Y^MD_\sigma Y_M\Big]\;,
\end{split}    
\end{equation}
where $\mathring{X}^\mu=\del_\tau X^\mu-u\,\del_\sigma X^\mu$. 
Since  \eqref{Tconsalmost} holds for arbitrary $\epsilon$ and $\alpha\,$, we obtain the energy-momentum conservation law in arbitrary gauge:
\begin{equation}\label{Tcons}
\begin{split}
\del_\tau\cH&=\del_\sigma(e\,\cN+u\,\cH)+\del_\sigma u\,\cH+\del_\sigma e\,\cN\;,\\
\del_\tau\cN&=\del_\sigma(e\,\cH+u\,\cN)+\del_\sigma u\,\cN+\del_\sigma e\,\cH\;.
\end{split}    
\end{equation}

Since diffeomorphism invariance can be used to fix the metric components $e$ and $u\,$, one is mostly interested in studying the model \eqref{finalaction} in conformal gauge, that 
corresponds to $e=1$ and $u=0\,$. 
The action \eqref{finalaction} then  reduces to
\begin{equation}\label{finalactionCG}
\begin{split}
S_{\rm c.g.}=&-\frac{1}{4\pi\alpha'}\int d^2\sigma\,\Big[g_{\mu\nu}\,\del^\alpha X^\mu\del_\alpha X^\nu+\epsilon^{\alpha\beta}\big(B_{\mu\nu}\,\del_\alpha X^\mu\del_\beta X^\nu-\cA_\mu{}^M\,D_\alpha Y_M\,\del_\beta X^\mu\big)\Big]  \\
&+\frac{1}{4\pi\alpha'}\,\int d^2\sigma\,\Big[D_\sigma Y^MD_\tau Y_M-\cH_{MN}\,D_\sigma Y^MD_\sigma Y^N\Big]\;.
\end{split}
\end{equation}
This has to be supplemented with the Virasoro constraints ${\cH=0}$, ${\cN=0}\,$. In conformal gauge it is useful to introduce light-cone worldsheet coordinates $\sigma^\pm:=\tau\pm\sigma\,$, for which one has
\begin{equation}
\del_\pm=\tfrac12\,(\del_\tau\pm\del_\sigma
)\;,\quad \eta_{+-}=-\frac12\;,\quad\eta^{+-}=-2\;,\quad \epsilon^{+-}=2\;,\quad\epsilon_{+-}=-\frac12\;.
\end{equation}
The Virasoro constraints can then be expressed in the more familiar form $T_{\pm\pm}=0\,$, defined by
\begin{equation}\label{Tpmpm}
\begin{split}
T_{++}&:=\pi\,(\cH_{\rm c.g.}+\cN_{c.g.})=\frac{1}{\alpha'}\,\Big(g_{\mu\nu}\,\del_+X^\mu\del_+X^\nu+\tfrac12\,\Pi^+_{MN}\,D_\sigma Y^MD_\sigma Y^N\Big)\;,\\
T_{--}&:=\pi\,(\cH_{\rm c.g.}-\cN_{c.g.})=\frac{1}{\alpha'}\,\Big(g_{\mu\nu}\,\del_-X^\mu\del_-X^\nu-\tfrac12\,\Pi^-_{MN}\,D_\sigma Y^MD_\sigma Y^N\Big)   \;,
\end{split}    
\end{equation}
where the $O(d,d)$ projectors are as in (\ref{diagonalHs}) and $D_\sigma=D_+-D_-\,$.

Conformal symmetry of the gauge fixed action \eqref{finalactionCG} is easily established as the global remnant of diffeomorphisms that preserve the conformal gauge choice.
From the transformation law \eqref{deltaHeu} one has the conditions
\begin{equation}\label{confgauge}
\delta e\rvert_{\rm conf.}=\del_\tau\epsilon-\del_\sigma\alpha =0\;,\quad \delta u\rvert_{\rm conf.}=\del_\tau\alpha-\del_\sigma\epsilon=0\;, \end{equation}
to preserve the conformal gauge. Since in this gauge Hamiltonian and Lagrangian parameters coincide, $\xi^\alpha=(\epsilon,\alpha)\,$, one can easily see that the conditions \eqref{confgauge} are equivalent to the usual analyticity 
\begin{equation}
\del_-\xi^+=0\;,\quad \del_+\xi^-=0\;.    
\end{equation}
The diffeomorphism transformations \eqref{diff2trans} give directly the conformal transformations leaving \eqref{finalactionCG} invariant:
\begin{equation}\label{conftransf}
\begin{split}
\delta_{\rm conf}X^\mu&=\xi^+\del_+X^\mu+\xi^-\del_-X^\mu\;,\\
\delta_{\rm conf}Y^M&=\xi^+\del_+Y^M+\xi^-\del_-Y^M-(\xi^++\xi^-)\Big[\Pi_-^M{}_N\,D_+Y^N+\Pi_+^M{}_N\,D_-Y^N\Big]   \;, 
\end{split}
\end{equation}
for analytic parameters $\xi^+(\sigma^+)\,$, $\xi^-(\sigma^-)\,$. Moreover, the usual $\frak{diff}_2$ algebra \eqref{diff2algebra} ensures that conformal transformations form two commuting copies of the classical Virasoro (Witt) algebra.

As a final remark, we notice that the conservation law \eqref{Tcons} reduces in conformal gauge to analyticity of the energy-momentum tensor:
\begin{equation}
\del_-T_{++}=0\;,\quad \del_+T_{--}=0\;,    
\end{equation}
despite the non-standard contributions from the $Y$ sector. In fact, we shall also notice that the self-duality relation \eqref{dualityarbitrarygauge}, that is $\cD^M=0\,$, can be written in conformal gauge as 
\begin{equation}\label{chiralityproj}
\Pi^\pm_{MN}D_\mp Y^N=0 \;. 
\end{equation}
If \eqref{chiralityproj} is imposed, both the conformal transformations of $Y^M$ and the $Y$ contribution to the stress-energy tensor assume the standard form
\begin{equation}
\begin{split}
&\delta_{\rm conf.}Y^M=\xi^+\del_+Y^M+\xi^-\del_-Y^M\;,\\
&T_{\pm\pm}=\frac{1}{\alpha'}\,\Big(g_{\mu\nu}\,\del_\pm X^\mu\del_\pm X^\nu+\tfrac12\,\cH_{MN}\,D_\pm Y^MD_\pm Y^N\Big)\;,
\end{split}
\end{equation}
but one should always keep in mind that \eqref{chiralityproj} is not a variational equation.

\section{Cancellation of Gravitational Anomalies}
In this appendix we verify explicitly that for the model carrying $d$ left-moving and $d$ right-moving chiral bosons the gravitational anomalies cancel. 
To this end we compute  the one-loop effective action for the gravitational field and  establish that gravitational anomalies can be canceled without spoiling $O(d,d)$ invariance.
For simplicity we consider the theory defined by the action \eqref{finalaction} for the case of vanishing gauge fields, $\cA_\mu{}^M=0$, and constant generalized metric: $\del_\mu\cH_{MN}=0\,$. The $Y$-sector decouples from the $X$-sector and reduces to the sum of left and right Floreanini-Jackiw actions coupled to gravity \cite{Henneaux:1987hz,Sonnenschein:1988ug,Bastianelli:1989cu}:
\begin{equation}\label{FJgravity}
S[Y,e_\pm]=\frac{1}{4\pi\alpha'}\,\int d^2\sigma\Big[\del_\sigma Y_L^M\,(\del_\tau-e_+\del_\sigma)Y_{L\,M}+\del_\sigma Y_R^M\,(\del_\tau-e_-\del_\sigma)Y_{R\,M}\Big]\;,    
\end{equation}
where we used the $O(d,d)$ projectors to define
\begin{equation}
Y^M_L:=\Pi_+^M{}_N\,Y^N\;,\quad Y^M_R:=\Pi_-^M{}_N\,Y^N   
\end{equation}
and introduced $e_\pm:=u\pm e\,$. The action \eqref{FJgravity} is invariant under two-dimensional diffeomorphisms acting as
\begin{equation}
\begin{split}
&\delta_\varepsilon Y^M_L=\varepsilon_+\,\del_\sigma Y^M_L\;,\qquad \delta_\varepsilon Y^M_R=\varepsilon_-\,\del_\sigma Y^M_R\;,\\
&\delta_\varepsilon e_\pm=\del_\tau\varepsilon_\pm-e_\pm\,\del_\sigma\varepsilon_\pm+\varepsilon_\pm\,\del_\sigma e_\pm\;,
\end{split}    
\end{equation}
where the parameters $\varepsilon_\pm=\alpha\pm\epsilon$ are given by
\begin{equation}
\varepsilon_\pm=\xi^\sigma+e_\pm\,\xi^\tau    
\end{equation}
in terms of the usual vector field $\xi^\alpha\,$.
The action is also invariant under two separate zero-mode local symmetries:
\begin{equation}\label{zeromodesymagain}
\delta_\Xi Y^M_L(\sigma,\tau)=\Xi^M_L(\tau)\;,\quad \delta_\Xi Y^M_R(\sigma,\tau)=\Xi^M_R(\tau)  \;. 
\end{equation}

In order to compute  the one-loop effective action for the gravitational field
we start by shifting the gravity fields as
\begin{equation}
e_+=1+\varphi_+\;,\quad e_-=-1-\varphi_-\;,
\end{equation}
so that $\varphi_\pm=0$ in conformal gauge. The action \eqref{FJgravity} then splits into a quadratic part and an interaction term, allowing for a well-defined perturbative treatment:
\begin{equation}
\begin{split}
S[Y,\varphi_\pm]&=\frac{1}{4\pi\alpha'}\int d^2\sigma\,\Big[\del_\sigma Y_L^M\,(\del_\tau-\del_\sigma)Y_{L\,M}+\del_\sigma Y_R^M\,(\del_\tau+\del_\sigma)Y_{R\,M}\Big]\\
&+\frac{1}{4\pi\alpha'}\int d^2\sigma\,\Big[\varphi_-\,\del_\sigma Y_R\cdot\del_\sigma Y_R-\varphi_+\,\del_\sigma Y_L\cdot \del_\sigma Y_L\Big]\\
&= S_L[Y_L,\varphi_+]+S_R[Y_R,\varphi_-]\;,
\end{split}    
\end{equation}
where the dot denotes contraction of $O(d,d)$ indices with $\eta_{MN}\,$.
Since the action is the sum of independent left and right terms, the path integral\footnote{We fix the normalization $Z$ to be the free $Y$-path integral, so that $W[0]=0\,$.} factorizes:
\begin{equation}
e^{i\,W[\varphi_\pm]}:=Z^{-1}\int DY_LDY_R\,e^{i\,S[Y,\varphi_\pm]}=e^{i\,W_L[\varphi_+]}e^{i\,W_R[\varphi_-]}\;. \end{equation}
We will thus focus on the left part of the effective action $W_L[\varphi_+]\,$, that can be written as the quantum average
\begin{equation}
e^{i\,W_L[\varphi_+]}=Z_L^{-1}\int DY_L\,e^{i\,S_L[Y_L,\varphi_+]}=:\left\langle e^{-\frac{i}{4\pi\alpha'}\int d^2\sigma\,\varphi_+\,\del_\sigma Y_L\cdot\del_\sigma Y_L}\right\rangle \;,   
\end{equation}
and just present the result for $W_R[\varphi_-]\,$.

Due to the zero-mode symmetry \eqref{zeromodesymagain}, the kinetic operator $\del_\sigma(\del_\tau-\del_\sigma)$ is not invertible. Gauge fixing \eqref{zeromodesymagain} with appropriate boundary conditions at asymptotic times \cite{Henneaux:1987hz} yields a trivial path integral for the zero-mode $Y_{L\,0}^M(\tau)\,$. The above path integral is thus understood as $\int D\widebar Y_L$ over the non-zero mode part of $Y^M_L$ only:
\begin{equation}
\widebar Y^M_L(\sigma,\tau):=Y^M_L(\sigma,\tau)-Y^M_{L\,0}(\tau)\;,  \end{equation}
whose propagator is well-defined and given by
\begin{equation}\label{propYY}
\left\langle\widebar Y^M_L(x)\,\widebar Y^N_L(y)\right\rangle = -4\pi\alpha'\,i\,\Pi^{MN}_+\int\frac{[d^2k]}{(2\pi)^2}\,e^{ik\cdot(x-y)}\,\frac{k_+}{k_1}\,\frac{1}{k^2-i\epsilon}\;,  
\end{equation}
where the momentum $k_\alpha:=(\omega,n)\,$, with discrete $n\in\mathbb{Z}$ in the $\sigma$-direction, and the ``integration'' measure is defined by
\begin{equation}\label{intmeasure}
\int[d^2k]:=\int_{-\infty}^{+\infty} d\omega\,\sum_{n\neq 0}\;.    
\end{equation}
We have also changed notation by denoting the worldsheet coordinates as $x^\alpha=(\tau,\sigma)\,$, in order not to confuse $\sigma^\alpha$ with the spatial component $\sigma\,$.

Equipped with the propagator \eqref{propYY} we can compute $W_L[\varphi_+]$ up to quadratic order:
\begin{equation}
W_L[\varphi_+]=id\,\Delta\,\int d^2x\,\varphi_+(x)-id\,\int d^2x\int\frac{d^2y}{(2\pi)^2}\frac{d^2p}{(2\pi)^2}\,e^{ip\cdot(x-y)}\varphi_+(x)\varphi_+(y)\,I(p)\;,  
\end{equation}
with the tadpole $\Delta$ and the bubble diagram $I(p)$ defined by
\begin{equation}
\Delta:=\int\frac{[d^2k]}{(2\pi)^2}\,\frac{k_1k_+}{k^2-i\epsilon}\;,\quad I(p):=\int[d^2k]\,\frac{k_1k_+}{k^2-i\epsilon}\,\frac{(p_1+k_1)(p_++k_+)}{(p+k)^2-i\epsilon} \;.   
\end{equation}
We regulate the frequency integrals (see \eqref{intmeasure}) by inserting a factor of $e^{i\epsilon'\omega}\,$. This is sufficient to make $I(p)$ finite, while a divergent contribution has still to be subtracted from the zero-point energy $\Delta\,$. Although the regulator manifestly breaks Lorentz symmetry, we will be only interested in the non-local part of the effective action, that is not affected by changing the regularization scheme. The finite result for the above diagrams are 
\begin{equation}
\Delta=-\frac{i}{48\pi}\;,\quad I(p)=-\frac{i\pi}{24}\,\frac{p_1^3-p_1}{p_-}    
\end{equation}
which allows us to write the effective action as
\begin{equation}\label{WLeft}
\begin{split}
W_L[\varphi_+]&=\frac{d}{48\pi}\int d^2x\,\varphi_++\frac{d}{96\pi}\int d^2x\,\varphi_+\left(\frac{\del_1^3+\del_1}{\del_-}\right)\varphi_++{\cal O}(\varphi_+^3)\\
&=\frac{d}{96\pi}\int d^2x\,\varphi_+\left(\frac{\del_+^3+\del_+}{\del_-}\right)\varphi_++{\rm local\;terms}+{\cal O}(\varphi_+^3)\;.
\end{split}
\end{equation}
Genuine anomalies are the ones that cannot be canceled by adding local counterterms (that reflect different regularizations) to the effective action. That is why we only need to focus on the non-local part of $W_L$ above.

We can now compute the gauge transformation of the effective action. 
Using the transformation law
\begin{equation}
\delta_\varepsilon\varphi_+=2\,\del_-\varepsilon_+-\varphi_+\,\del_1\varepsilon_++\varepsilon_+\,\del_1\varphi_+  \;,  
\end{equation}
it is easy to see that the anomalous variation of \eqref{WLeft} is given by
\begin{equation}
\delta_\varepsilon W_L^{\rm n.l.}[\varphi_+]=-\frac{d}{24\pi}\int d^2x\,\varepsilon_+(\del_+^3+\del_+)\varphi_++{\cal O}(\varphi_+^2)\;.    
\end{equation}
The term $\varepsilon_+\,\del_+\varphi_+$ can be canceled by adding a local counterterm proportional to ${\varphi_+-\tfrac12\,\varphi_+^2}\,$, but the term cubic in $\del_+$ cannot be canceled and represents the genuine gravitational anomaly (to lowest order in $\varphi_+$) of the chiral bosons $Y^M_L$ on the cylinder.
The anomaly is the same obtained on the plane from $d$ left-moving Floreanini-Jackiw bosons \cite{Sonnenschein:1988ug,Bastianelli:1990ev,Giaccari:2008zx}. This ensures that adding the contribution $W_R[\varphi_-]$ allows to cancel the gravitational anomaly completely. 

To be more explicit, adding the contribution from the right-moving fields $Y^M_R$ one obtains the full non-local contribution:
\begin{equation}\label{Wnonloc}
W^{\rm n.l.}[\varphi_\pm] = \frac{d}{96\pi}\int d^2x\,\Big\{\varphi_+\left(\frac{\del_+^3+\del_+}{\del_-}\right)\varphi_++\varphi_-\left(\frac{\del_-^3+\del_-}{\del_+}\right)\varphi_- \Big\}+{\cal O}(\varphi_\pm^3) \;.
\end{equation}
The gravitational anomaly of the above expression can indeed be canceled by adding a local counterterm that involves the third degree of freedom of the worldsheet metric: the conformal factor $\Omega$. Denoting the deviation of $\Omega$ from its flat space value by $\phi:=\Omega-1$ 
one can add to 
\eqref{Wnonloc} a local counterterm $\Delta W[\varphi_\pm,\phi]$ and define the effective action as
\begin{equation}\label{Weff}
\begin{split}
W_{\rm eff}[\varphi_\pm,\phi]&=\frac{d}{96\pi}\int d^2x\,\Big\{\varphi_+\,\frac{\del_+^3}{\del_-}\,\varphi_++\varphi_-\,\frac{\del_-^3}{\del_+}\,\varphi_-+2\,\varphi_+\,\del_+\del_-\varphi_-\\
&\hspace{25mm}-4\,\phi\,\big(\del_+^2\varphi_++\del_-^2\varphi_-\big)+4\,\phi\,\del_+\del_-\phi \Big\}\\
&+\frac{d}{96\pi}\int d^2x\,\Big\{\varphi_+\,\frac{\del_+}{\del_-}\,\varphi_++2\,\varphi_+-\varphi_+^2+\varphi_-\,\frac{\del_-}{\del_+}\,\varphi_-+2\,\varphi_--\varphi_-^2\Big\}+{\cal O}(\rm fields^3)\;.
\end{split}    
\end{equation}
One can check that the above effective action is invariant under diffeomorphisms (to lowest order in the fields, since we are considering only the quadratic part of $W_{\rm eff}$) with transformations
\begin{equation}
\begin{split}
&\delta_\varepsilon\varphi_+=2\,\del_-\varepsilon_+-\varphi_+\,\del_1\varepsilon_++\varepsilon_+\,\del_1\varphi_+\;,\quad  \delta_\varepsilon\varphi_-=-2\,\del_+\varepsilon_--\varphi_-\,\del_1\varepsilon_-+\varepsilon_-\,\del_1\varphi_-\;,\\
&\delta\phi=\del_+\varepsilon_+-\del_-\varepsilon_-+\tfrac12\,\del_+[\varphi_+(\varepsilon_--\varepsilon_+)+2\,\phi\,\varepsilon_+]+\tfrac12\,\del_-[\varphi_-(\varepsilon_--\varepsilon_+)-2\,\phi\,\varepsilon_-]+{\cal O}({\rm fields^2})\;.
\end{split}    
\end{equation}
Let us mention that the last line in \eqref{Weff} is invariant by itself and the tadpoles (linear terms) in $\varphi_\pm$ just reflect the non-zero Casimir energy on the cylinder, \emph{i.e.} $\langle T_{\pm\pm}\rangle\neq0\,$. 

As it happens for ordinary scalars, the price to pay to restore diffeomorphism invariance is the breakdown of Weyl symmetry. While $\varphi_\pm$ are exactly Weyl invariant, $\delta_\omega\phi=2\,\omega+{\cal O}(\phi)$
and one readily obtains
\begin{equation}
\delta_\omega W_{\rm eff}[\varphi_\pm,\phi]=\frac{d}{12\pi}\int d^2x\,\omega\,(2\,\del_+\del_-\phi-\del_+^2\varphi_+-\del_-^2\varphi_-) +{\cal O}({\rm fields^2})\;.   
\end{equation}
To lowest order in the fields this is
\begin{equation}
\delta_\omega W_{\rm eff}[\varphi_\pm,\phi]=\frac{d}{24\pi}\int d^2x\,\sqrt{-h}\,\omega\,R \;, 
\end{equation}
thus yielding the trace anomaly
\begin{equation}
\langle T^\alpha{}_\alpha\rangle=-\frac{d}{12}\,R   . 
\end{equation}
This confirms that the internal $Y^M$ sector just contributes to the trace anomaly with $d$ units of both left and right central charge. Indeed, including the $n$ external ordinary bosons $X^\mu\,$, one has the requirement $n+d=26$ for criticality.


\bibliography{ODDStrings.bib}
\bibliographystyle{unsrt}



\end{document}